\documentclass[reprint,aps,prl,superscriptaddress]{revtex4-2}

\usepackage{amsmath,amssymb,amsfonts,amsthm} 
\usepackage{graphicx}
\usepackage{dcolumn}
\usepackage{bm}
\usepackage[bookmarks=true,colorlinks=true,citecolor=cyan,linkcolor=cyan,filecolor=cyan,urlcolor=cyan]{hyperref} 
\usepackage{braket} 
\usepackage{tikz} 
\usepackage{comment} 
\usepackage[normalem]{ulem} 
\usepackage{booktabs}
\usepackage{algorithm} 
\usepackage{mathtools}
\usepackage{orcidlink}

\DeclareMathOperator{\Tr}{Tr}

\theoremstyle{plain}
\newtheorem{thm}{Theorem}
\newtheorem{prop}{Proposition}

\newtheorem{lem}{Lemma}

\theoremstyle{definition}

\begin{document}

\title{Virtual Autopurification in Translationally Invariant Quantum States}

\author{Atsushi Iwaki\,\orcidlink{0000-0002-5345-8045}}
\email{atsushi-iwaki@phys.s.u-tokyo.ac.jp}
\affiliation{Department of Physics, University of Tokyo, 7-3-1 Hongo, Bunkyo-ku, Tokyo 113-0033, Japan}

\author{Hideaki Hakoshima\,\orcidlink{0000-0003-1295-9801}}
\affiliation{Graduate School of Engineering Science, University of Osaka, 1-3 Machikaneyama, Toyonaka, Osaka 560-8531, Japan}
\affiliation{Center for Quantum Information and Quantum Biology, University of Osaka, 1-2 Machikaneyama, Toyonaka, Osaka 560-0043, Japan}

\author{Nobuyuki Yoshioka\,\orcidlink{0000-0001-6094-8635}}
\affiliation{International Center for Elementary Particle Physics, University of Tokyo, 7-3-1 Hongo, Bunkyo-ku, Tokyo 113-0033, Japan}

\date{\today}

\begin{abstract}
We introduce virtual autopurification (VAP), 
a protocol for virtual purification that operates entirely 
within a single translationally invariant system. 
Virtual purification suppresses thermal fluctuations and decoherence noise 
by amplifying the dominant pure component of a mixed state, 
but its standard implementation requires multiple copies of the state 
and global entangled measurements. 
VAP avoids the multiple-copy overhead by constructing virtual copies from spatially separated subregions 
and performing entangled measurements only between them. 
Assuming clustering of correlations, 
we show that VAP reproduces the expectation values of the standard virtual purification protocol, 
with discrepancies that decay exponentially 
in both the buffer length and the inter-subregion distance. 
Numerical simulations support the theoretical predictions 
and demonstrate accurate implementation of virtual cooling with reduced resource requirements. 
Our results establish VAP as a resource-efficient and scalable approach 
to quantum-state purification in near-term devices.
\end{abstract}

\maketitle

\emph{\textbf{Introduction.---}}
Among digital quantum simulation tasks, thermal state preparation is a central application of quantum computing 
with potential impact across condensed-matter physics, quantum chemistry, 
and high-energy physics.
The goal is to prepare a Gibbs state $\rho_\beta \propto e^{-\beta H}$ 
for a given Hamiltonian $H$ and temperature $T=1/\beta$ on quantum devices, 
which is essential for faithfully reproducing realistic experimental conditions. 
Despite substantial theoretical progress 
\cite{Brando2018, Chen2025, Rouze2026-1, Rouze2026-2} 
and experimental demonstrations 
\cite{Wu2019, Motta2019, Zhu2020, Sagastizabal2021}, 
reaching lower temperatures is still difficult, 
because thermal fluctuations and decoherence must be suppressed strongly. 

\begin{figure}
    \centering
    \includegraphics[width=\hsize]{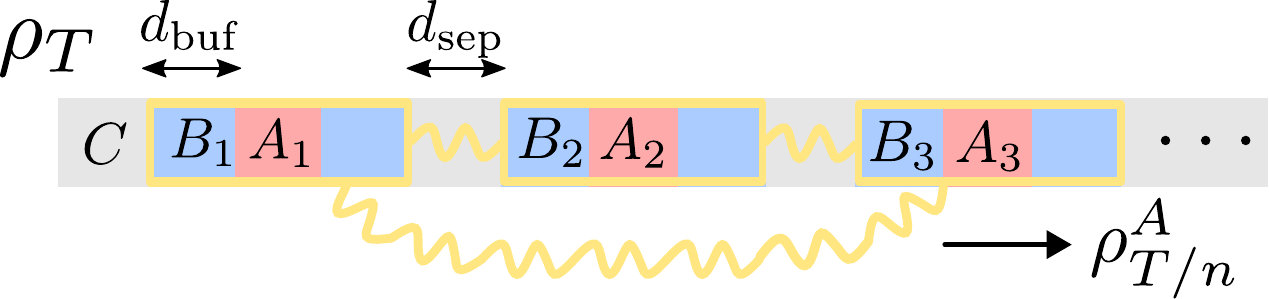}
    \caption{
    Schematic illustration of the VAP protocol. 
    Identical subregions $\{A_i B_i\}_{i=1}^n$ are chosen 
    in a translationally invariant Gibbs state $\rho_T$, 
    with $A$ supporting the observable and $B$ providing a correlation buffer. 
    Entangled measurements between distant subregions realize virtual cooling to temperature $T/n$. 
    }
    \label{fig:schematic}
\end{figure}

A promising route to alleviating these difficulties is virtual purification, 
a post-processing technique that extracts low-temperature observables 
from noisy or higher-temperature states 
\cite{Cotler2019, Koczor2021, Huggins2021, OBrien2023, Cai2021, Czarnik2021}. 
By amplifying the dominant pure component of a mixed state using powers of $\rho$ as $\rho^n$ via swap techniques, 
virtual purification suppresses thermal noise and decoherence. 
For a Gibbs state $\rho_\beta$, 
this amounts to a virtual cooling $\rho_\beta \mapsto \rho_{n\beta}$, 
increasing the inverse temperature by a factor of $n$.
However, 
the standard formulation, or \emph{fully virtual purification} (FVP), 
requires globally entangled measurements over multiple physical copies, 
and moreover incurs exponential sampling overhead with system size \cite{Vikstl2024}.

A first step toward reducing such an overhead was provided 
by \emph{localized virtual purification} (LVP) \cite{Hakoshima2024}, 
which exploits the exponential clustering property in geometrically local systems 
to replace the globally entangled measurement 
with the one supported on the vicinity of the target observable of interest. 
While such a technique lifts the need for exponential sampling overhead, 
it still leaves the common crucial property in condensed matter problems: translational symmetry.
Under clustered correlations, 
each local window of support can be expected to exhibit nearly identical reduced density matrices, 
while it has remained theoretically unclear whether it can be used 
as an operational resource for virtual purification. 

In this work, 
we formulate this idea as \emph{virtual autopurification} (VAP), 
a copy-free virtual purification protocol for translationally invariant quantum states, 
and prove its validity by providing a performance guarantee. 
As illustrated in Fig.~\ref{fig:schematic}, 
spatially separated subregions in a translationally invariant system 
act as approximately independent copies. 
Assuming exponential clustering, 
we show that VAP reproduces  expectation values of the standard purification
up to a bias that decays exponentially in both the buffer length 
and the inter-subregion separation, 
while circumventing the need for exponential overhead 
with respect to the system size (see Table~\ref{tab:protocol-comparison}).
Numerical simulations corroborate these bounds 
and demonstrate accurate virtual cooling with substantially reduced resources, 
establishing VAP as a scalable route to virtual purification.

\begin{table}[t]
    \caption{
    Comparison of fully virtual purification (FVP),
    localized virtual purification (LVP),
    and virtual autopurification (VAP).
    Here, $n$ is an index for a  virtually purified state $\rho^n$ 
    and $N$ is the total number of lattice sites. 
    Observable clustering and trace-norm clustering 
    are given by Eqs.~\eqref{eq:exp-clustering} 
    and \eqref{eq:clustering_state}, respectively.
    }
    \label{tab:protocol-comparison}
    \centering
    \small
    \setlength{\tabcolsep}{3pt}
    \begin{tabular}{@{}lccc@{}}
        \toprule
        & FVP & LVP & VAP (ours) \\
        \midrule
        \# copies per shot
          & $n$ & $n$ & $1$ \\
        Sampling overhead
          & $\exp\mathcal{O}(nN)$ 
          & $\exp\mathcal{O}(n)$ 
          & $\exp\mathcal{O}(n)$ \\
        Clustering assumption
          & none
          & \begin{tabular}[c]{@{}c@{}}
                Observable \\ clustering
            \end{tabular}
          & \begin{tabular}[c]{@{}c@{}}
                Trace-norm \\ clustering
            \end{tabular} \\
        \bottomrule
    \end{tabular}
\end{table}

\emph{\textbf{Virtual purification.---}}
We begin by briefly reviewing the existing purification protocols 
for simulating the Gibbs state of a geometrically-local Hamiltonian $H$.
Given a quantum state $\rho$ and an integer $n$, 
FVP estimates an observable $o_A$ 
supported on a local region $A$ 
via $\langle o_A \rangle_{\mathrm{FVP}} = \Tr[\rho^n \, o_A] / \Tr[\rho^n]$
as introduced in Refs.~\cite{Cotler2019, Koczor2021, Huggins2021}, 
which is designed to amplify the dominant component of $\rho$.
To implement FVP on a quantum device, 
one uses the cyclic shift operator $D_n$ acting on $n$ copies, which gives 
$\langle o_A \rangle_{\mathrm{FVP}}
= \Tr[\rho^{\otimes n} \, D_n \, o_A]
/ \Tr[\rho^{\otimes n} \, D_n]$. 
FVP therefore requires globally entangled operation for measuring
$D_n \, o_A$ and $D_n$, 
which can be performed, for example, via Hadamard tests \cite{Ekert2002}. 

LVP reduces these measurement resources by restricting the cyclic shift 
to a local subsystem $AB$, where $B$ is a buffer surrounding $A$ 
and $C$ is the remainder of the system \cite{Hakoshima2024}. 
The resulting estimator is 
$\langle o_A \rangle_{\mathrm{LVP}}
= \Tr[(\rho_{AB})^n \, o_A]/\Tr[(\rho_{AB})^n]$, 
where $\rho_{AB}=\Tr_C[\rho]$. 
Its accuracy follows from \textit{observable clustering}: 
for normalized operators $M_R$ and $M_{R'}$ supported on disjoint regions,
\begin{equation}
    |\Tr[\rho \, M_R M_{R'}]
      -\Tr[\rho \, M_R]\Tr[\rho \, M_{R'}]|
    \le \mathcal{O}(e^{-d_{RR'}}),
    \label{eq:exp-clustering}
\end{equation}
where $d_{RR'}$ is the distance between the regions. 
This property holds, in particular, for Gibbs states of local Hamiltonians 
at sufficiently high temperatures
\cite{Araki1969, Park1995, Ueltschi2004, Kliesch2014, Frhlich2015}. 
Under this condition, the FVP--LVP error satisfies
\begin{equation}
    \left|\langle o_A \rangle_{\mathrm{FVP}}
    -\langle o_A \rangle_{\mathrm{LVP}}\right|
    \le \mathcal{O}(e^{-d_\mathrm{buf}}),
    \label{eq:FVP-LVP-bound}
\end{equation}
where $d_\mathrm{buf}$ is the distance between $A$ and $C$, 
as shown in Fig.~\ref{fig:schematic}. 
Thus, LVP reproduces FVP up to exponentially small corrections 
while avoiding global measurements and substantially reducing 
the sampling overhead.

\emph{\textbf{Virtual autopurification.---}}
The essential idea of VAP is 
to eliminate the multiple-copy requirement of LVP
by constructing virtual copies 
within a single translationally invariant quantum state. 
When correlations decay exponentially, 
two spatially separated regions behave approximately independently; 
thus the LVP estimator can be reproduced by performing an entangled measurement 
between distant, isomorphic subregions of a single copy, 
as illustrated in Fig.~\ref{fig:schematic}. 

For clarity, we present the analytic calculation for the case $n=2$. 
Let $AB$ be the subsystem used in LVP and $A'B'$ be a translated copy of $AB$ 
separated by an inter-copy distance $d_\mathrm{sep}$. 
The remainder of the system is denoted by $C$. 
The VAP estimator is defined as
\begin{equation}
    \langle o_A \rangle_\mathrm{VAP}
    = \frac{\Tr[\rho_{ABA'B'} \, S_{AB,A'B'} \, o_A]}
           {\Tr[\rho_{ABA'B'} \, S_{AB,A'B'}]},
    \label{eq:VAP-def}
\end{equation}
where $S_{AB,A'B'}$ swaps the two subsystems $AB$ and $A'B'$. 
Assuming that the state is translationally invariant as $\rho_{AB} = \rho_{A'B'}$, 
the LVP estimator can be rewritten as
\begin{equation}
    \langle o_A \rangle_\mathrm{LVP}
    = \frac{\Tr[\rho_{AB} \otimes \rho_{A'B'} \, S_{AB,A'B'} \, o_A]}
           {\Tr[\rho_{AB} \otimes \rho_{A'B'} \, S_{AB,A'B'}]}.
    \label{eq:LVP-rewrite}
\end{equation}
The difference between the two estimators is given by
\begin{align}
    &\langle o_A \rangle_\mathrm{VAP}
      - \langle o_A \rangle_\mathrm{LVP} \notag \\
    &= \frac{\Tr[(\rho_{ABA'B'} - \rho_{AB} \otimes \rho_{A'B'})\,
         S_{AB,A'B'} \, \tilde{o}_A]}
            {\Tr[(\rho_{AB})^2]},
    \label{eq:VAP-LVP}
\end{align}
where $\tilde{o}_A = o_A - \langle o_A \rangle_\mathrm{VAP}$. 

If the quantum state on $AB$ and $A'B'$ were completely separated, 
so that $\rho_{ABA'B'} = \rho_{AB} \otimes \rho_{A'B'}$,
the LVP and VAP estimators would coincide.
In general, however, this factorization does not hold exactly.
We therefore assume exponential clustering 
at the level of reduced density operators, 
which we refer to as \textit{trace-norm clustering}:
\begin{align}
    \| \rho_{R R'} - \rho_{R} \otimes \rho_{R'} \|_1 \le \mathcal{O}(e^{-d_{RR'}}).
    \label{eq:clustering_state}
\end{align}
This is a stronger assumption than the clustering of observables 
in Eq.~\eqref{eq:exp-clustering}. 
This assumption exactly holds for one-dimensional systems 
at arbitrary temperature \cite{Bluhm2022}. 
In higher-dimensional systems, 
it is also known to hold under the assumption 
that the effective Hamiltonians associated with local reduced density operators 
of the Gibbs state remain local \cite{Bluhm2025}. 

Building on this clustering property, 
we analyze the VAP--LVP discrepancy through second order in $\lambda$,
a parameter that scales the boundary couplings 
between the virtual-copy regions and the remainder of the system.
Concretely, 
together with the high-temperature condition 
specified in Lemma~1 of the Supplemental Material (SM) \cite{SM}, 
we obtain the following bound.

\begin{thm} \label{thm:FVP-VAP} 
Assuming trace-norm clustering 
and sufficiently high temperature 
indicating quasi-locality 
of the imaginary-time evolution, 
the difference between the FVP and VAP estimators
is suppressed with both the buffer length $d_\mathrm{buf}$ 
and the inter-copy separation $d_\mathrm{sep}$ as 
\begin{equation}
    |\langle o_A \rangle_{\mathrm{FVP}}-\langle o_A \rangle_{\mathrm{VAP}}|
    \le \mathcal{O}(e^{-d_\mathrm{buf}}) 
    + \mathcal{O}(\lambda^2 e^{-d_\mathrm{sep}})
    + \mathcal{O}(\lambda^3).
\end{equation}
\end{thm}

This theorem follows by combining the previously established bound
between FVP and LVP in Eq.~\eqref{eq:FVP-LVP-bound}
with the proposition below.

\begin{prop} \label{prop:VAP-LVP}
Under the assumptions stated in Theorem~\ref{thm:FVP-VAP}, 
the difference between the LVP and VAP estimators 
decays exponentially with the inter-copy distance:
\begin{equation}
    |\langle o_A \rangle_{\mathrm{LVP}} - \langle o_A \rangle_{\mathrm{VAP}}|
    \le \mathcal{O}(\lambda^2 e^{-d_\mathrm{sep}})
    + \mathcal{O}(\lambda^3).
\end{equation}
\end{prop}

\emph{Proof sketch.}
We compare the two density operators 
$\rho_{ABA'B'}$ and $\rho_{AB} \otimes \rho_{A'B'}$ in Eq.~\eqref{eq:VAP-LVP}
perturbatively around a decoupled Hamiltonian. 
Let $H=H_0+V$, where $H_0$ contains the interactions within $AB$, $A'B'$, 
and $C$, while $V=V_{BC}+V_{B'C}$ contains the boundary couplings to $C$.
By introducing a coupling parameter
$\lambda$ for $V$, we consider the perturbed thermal states
$\rho(\lambda) \propto e^{-\beta (H_0 + \lambda V)}$.
The VAP and LVP estimators coincide up to first order in $\lambda$, 
while the second-order deviation of the density operators is given by 
\begin{align}
    &\rho_{ABA'B'} - \rho_{AB} \otimes \rho_{A'B'} \notag \\
    &= \lambda^2 \rho_{AB}^{(0)} \otimes \rho_{A'B'}^{(0)}
    \int_0^\beta ds \int_0^{s} dt\,
    \left[ \Delta_{BB'}(s,t) \right. \notag \\
    &\left. \quad - \Delta_B(s,t) - \Delta_{B'}(s,t) + \Delta(s,t) \right]
    + \mathcal{O}(\lambda^3).
    \label{eq:diff_pert}
\end{align}
Here, $\rho^{(0)}$ denotes the thermal state at $\lambda=0$, 
and $\Delta_{BB'}(s,t)$ encodes the two-copy correlations,
\begin{align}
    &\Delta_{BB'}(s,t)
    = \Tr_C[\rho_C^{(0)} \, V_{BC}^{\mathrm{I}}(s) V_{B'C}^{\mathrm{I}}(t)]
    - \tilde{V}_B(s)\tilde{V}_{B'}(t) \notag \\
    &\quad
    + \Tr_C[\rho_C^{(0)} \, V_{B'C}^{\mathrm{I}}(s) V_{BC}^{\mathrm{I}}(t)]
    - \tilde{V}_{B'}(s)\tilde{V}_{B}(t),
\end{align}
with $V_{BC}^{\mathrm{I}}(s) = e^{sH_0} \, V_{BC} \, e^{-sH_0}$ and
$\tilde{V}_B(s) = \Tr_C[\rho_C^{(0)} \, V_{BC}^{\mathrm{I}}(s)]$ denoting the corresponding
effective operator associated with $B$.
We further define the reduced operator
$\Delta_B(s,t) = \Tr_{A'B'}[\rho_{A'B'}^{(0)} \Delta_{BB'}(s,t)]$,
with $\Delta_{B'}(s,t)$ obtained analogously by tracing out $AB$.
The corresponding scalar function is defined as
$\Delta(s,t) = \Tr[\rho^{(0)} \Delta_{BB'}(s,t)]$. 
Substituting Eq.~\eqref{eq:diff_pert} into the expression for the difference 
between the VAP and LVP estimators in Eq.\eqref{eq:VAP-LVP}, 
we obtain four classes of terms. 
The scalar term $\Delta(s,t)$ vanishes because it multiplies 
the zeroth-order expectation value of the centered observable. 
For the terms involving $\Delta_B(s,t)$ and $\Delta_{B'}(s,t)$, 
Lemma~1 of SM \cite{SM} approximates 
the imaginary-time-evolved boundary interactions by local operators 
with exponentially small errors. 
The local-observable clustering condition in Eq.~\eqref{eq:exp-clustering} 
then gives an $\mathcal{O}(e^{-d_\mathrm{sep}})$ bound. 
The $\Delta_{BB'}(s,t)$ term retains correlations between reduced density operators 
even after the imaginary-time-evolved boundary interactions are replaced 
by their local approximations, and therefore requires the stronger condition 
in Eq.~\eqref{eq:clustering_state}, which gives the same exponential bound. 
Combining these estimates gives the exponential suppression stated 
in Proposition~\ref{prop:VAP-LVP}. 
The detailed derivation for $n=2$ is presented in SM \cite{SM}.
Since connected contributions at second order in $\lambda$
involve at most two virtual copies,
the same analysis applies to any fixed $n\ge3$.

Theorem~\ref{thm:FVP-VAP} bounds the systematic bias of VAP relative to FVP. 
We quantify the required resources by the space--time overhead, 
defined as the measurement cost multiplied by the number of copies required per shot. 
For estimating the expectation value of a Pauli operator $P_A$, 
the FVP variance is bounded as
\begin{align}
    \frac{1}{\Tr[\rho^n]^2}
    \le \mathrm{Var}(P_A)_\mathrm{FVP}
    \le \frac{1+\langle P_A\rangle^2_\mathrm{FVP}}{\Tr[\rho^n]^2}.
\end{align}
This motivates defining the measurement cost of FVP as $\Tr[\rho^n]^{-2}$, 
following the Supplemental Material of Ref.~\cite{Hakoshima2024}. 
By the same reasoning, 
the measurement costs of LVP and VAP are given 
by $\Tr[(\rho_{AB})^n]^{-2}$ and $|\Tr[\rho \, D_n^{AB}]|^{-2}$, respectively. 
Combining these measurement costs with the number of physical copies required per shot, 
we define the space--time overhead $\mathcal{C}_{\mathrm{st}}$ as 
$\mathcal{C}_{\mathrm{st}}^{\mathrm{FVP}} = n / \Tr[\rho^n]^2$, 
$\mathcal{C}_{\mathrm{st}}^{\mathrm{LVP}} = n / \Tr[(\rho_{AB})^n]^2$ and 
$\mathcal{C}_{\mathrm{st}}^{\mathrm{VAP}} = 1 / |\Tr[\rho \, D_n^{AB}]|^2$
where $D_n^{AB}$ denotes the cyclic shift among the $n$ spatially separated 
subregions used as virtual copies within a single system. 
Their leading scaling is 
\begin{align}
    \mathcal{C}_\mathrm{st}^{\mathrm{FVP}}
    &\sim n\exp\!\left[2(n-1)N s_\mathrm{th}\right], \notag\\
    \mathcal{C}_\mathrm{st}^{\mathrm{LVP}}
    &\sim n\exp\!\left[2(n-1)N_{AB}s_\mathrm{th}\right], \notag\\
    \mathcal{C}_\mathrm{st}^{\mathrm{VAP}}
    &\sim \exp\!\left[2(n-1)N_{AB}s_\mathrm{th}\right].
    \label{eq:overhead-scaling}
\end{align}
Here, $N$ and $N_{AB}$ are the numbers of sites in the full system 
and region $AB$, respectively, and $s_\mathrm{th}$ is the thermal entropy density.
Thus, LVP and VAP replace the dependence on the total system size $N$ 
with that on the local subsystem size $N_{AB}$, while VAP further removes 
the prefactor $n$ by requiring only one physical copy per shot. 

\emph{\textbf{Numerical simulations.—}}
We numerically test the theoretical predictions above and benchmark the performance of VAP.
As a benchmark, 
we consider the one-dimensional transverse-field Ising model 
with open boundary conditions,
\begin{equation}
    H = -\sum_{i=1}^{N-1} \sigma_i^z \sigma_{i+1}^z
    - g \sum_{i=1}^N \sigma_i^x ,
\end{equation}
where $\sigma_i^\alpha \, (\alpha = x, y, z)$ are Pauli operators at site $i$. 
In the following calculations, we focus on the critical point $g=1$. 
To access large system sizes and directly manipulate thermal mixed states, 
we employ matrix-product density operators (MPDOs) 
\cite{Verstraete2004, Zwolak2004, Feiguin2005}
implemented using the ITensor library \cite{itensor1, itensor2}.

\begin{figure}
    \centering
    \includegraphics[width=\hsize]{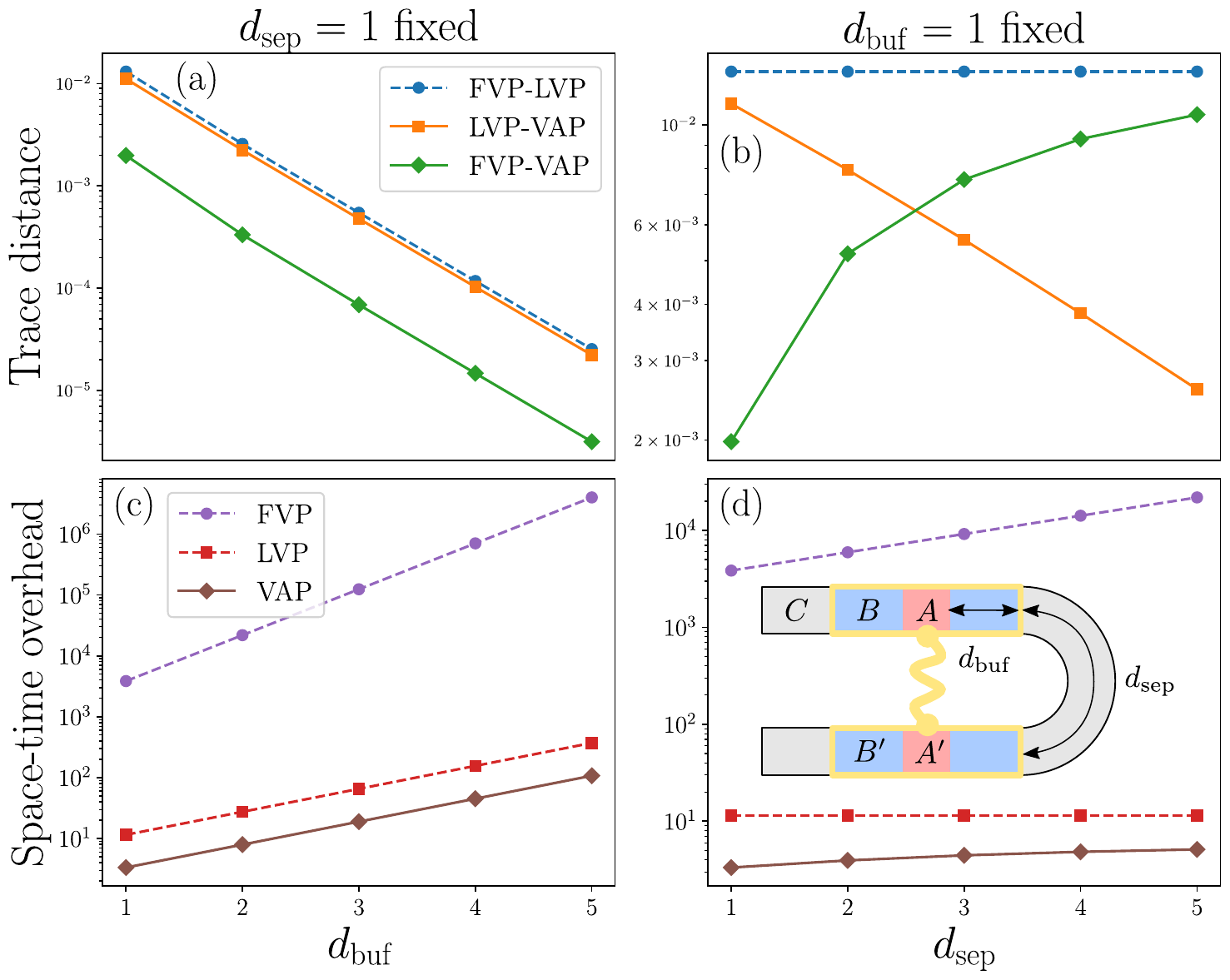}
    \caption{
    (a), (b) Trace distances between the effective reduced density operators 
    of FVP, LVP, and VAP as functions of $d_\mathrm{buf}$ and $d_\mathrm{sep}$, respectively.
    (c), (d) Corresponding space--time overheads $\mathcal{C}_\mathrm{st}$.
    The subsystem configuration is shown in the inset.
    We set $\beta=1$, $|A|=1$, and $|B|=2d_\mathrm{buf}$, 
    with five additional boundary sites at each end of the chain, giving
    $N=12+4d_\mathrm{buf}+d_\mathrm{sep}$.
    }
    \label{fig:tr_dist}
\end{figure}

Following the subsystem arrangement shown in Fig.~\ref{fig:tr_dist} (inset),
we study how the distances $d_\mathrm{buf}$ and $d_\mathrm{sep}$ control the
differences among the three estimators for $n=2$.
For each method, 
we define the effective reduced density operator on region $A$ as
$\rho_\mathrm{FVP}^A = \Tr_{\bar{A}}[\rho^2] / \Tr[\rho^2]$,
$\rho_\mathrm{LVP}^A = \Tr_{\bar{A}}[(\rho_{AB})^2] / \Tr[(\rho_{AB})^2]$ and 
$\rho_\mathrm{VAP}^A = \Tr_{\bar{A}}[S_{AB,A'B'} \, \rho] / \Tr[S_{AB,A'B'} \, \rho]$
where $\bar{A}$ is a complement of the subsystem $A$. 
We refer to these as effective reduced density operators because
$\rho_\mathrm{VAP}^A$ is generally non-Hermitian, 
while still reproducing the correct VAP expectation values.

Figure~\ref{fig:tr_dist} shows the trace distances 
between the effective reduced density operators 
and the corresponding space--time overheads for the three protocols, 
as functions of $d_\mathrm{buf}$ and $d_\mathrm{sep}$. 
The exponential decay of the FVP--LVP discrepancy with $d_\mathrm{buf}$ is consistent 
with the theoretical result established in Ref.~\cite{Hakoshima2024}. 
Likewise, the exponential decay of the LVP--VAP discrepancy with $d_\mathrm{sep}$ 
supports the validity of Proposition~\ref{prop:VAP-LVP}.
At the same time, 
the numerical results also suggest a stronger scaling than 
that stated in Proposition~\ref{prop:VAP-LVP}, 
since the LVP--VAP discrepancy appears to decay exponentially 
in both $d_\mathrm{buf}$ and $d_\mathrm{sep}$:
\begin{align}
    |\langle o_A \rangle_{\mathrm{LVP}}-\langle o_A \rangle_{\mathrm{VAP}}|
    =\mathcal{O}(e^{-d_\mathrm{buf}}) \times \mathcal{O}(e^{-d_\mathrm{sep}}).
\end{align}
Consequently, when $d_\mathrm{sep}$ is sufficiently large for the LVP--VAP
discrepancy to become negligible, the VAP approximation error relative to FVP
is governed primarily by $d_\mathrm{buf}$.
The space--time overhead further highlights the practical advantage of LVP and VAP. 
For fixed $d_\mathrm{sep}$, the suppression of the LVP 
overhead relative to FVP scales as
$\mathcal{C}_\mathrm{st}^{\mathrm{FVP}}/
\mathcal{C}_\mathrm{st}^{\mathrm{LVP}}
\sim\exp[2(n-1)N_Cs_\mathrm{th}]$,
where $N_C=N-N_{AB}$. 
Thus, the LVP overhead is exponentially smaller than the FVP overhead 
in the size of $C$. For fixed $d_\mathrm{buf}$, $N_{AB}$ remains constant 
as $d_\mathrm{sep}$ increases, explaining the nearly constant LVP and VAP 
overheads. VAP further reduces the overhead relative to LVP by approximately 
a factor of $n$, since it requires only one physical copy per shot.

\begin{figure}
    \centering
    \includegraphics[width=\hsize]{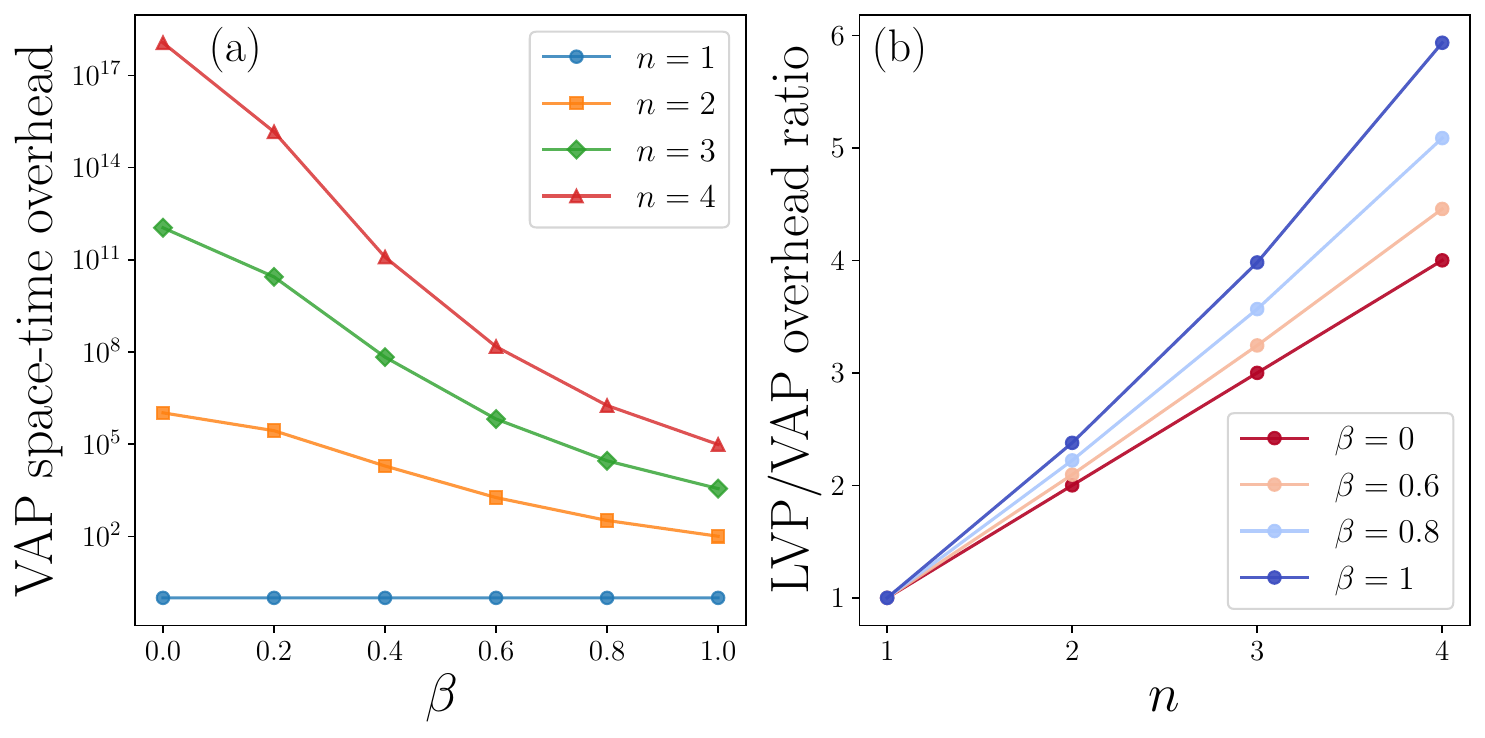}
    \caption{
    (a) Space--time overhead of the VAP protocol 
    $ \mathcal{C}_\mathrm{st}^{\mathrm{VAP}} $ 
    as a function of $\beta$ for several numbers of virtual copies $n$.
    (b) Ratio of the LVP overhead to the VAP overhead 
    $ \mathcal{C}_\mathrm{st}^{\mathrm{LVP}}/\mathcal{C}_\mathrm{st}^{\mathrm{VAP}} $
    as a function of $n$ for $\beta=0,\,0.6,\,0.8,$ and $1$.
    The calculations use $N=60$, $|A|=2$, and $d_\mathrm{buf}=4$.
    For each $n$, the virtual copies are arranged to maximize their 
    inter-copy separations; for $n=4$, this gives $d_\mathrm{sep}=4$.
    }
    \label{fig:overhead}
\end{figure}

Figure~\ref{fig:overhead} plots the space--time overhead 
for the number of virtual copies varied from $n=1$ to $4$. 
As shown in Fig.~\ref{fig:overhead}(a), 
the overhead of VAP decreases toward lower temperatures. 
This behavior is consistent with the entropy scaling in 
Eq.~\eqref{eq:overhead-scaling}: lowering the temperature reduces 
$s_\mathrm{th}$ and hence the space--time overhead of VAP.
Figure~\ref{fig:overhead}(b) shows 
$ \mathcal{C}_\mathrm{st}^{\mathrm{LVP}}/\mathcal{C}_\mathrm{st}^{\mathrm{VAP}} $, 
i.e., the advantage of VAP over LVP. 
In the high-temperature regime, 
the expected factor-of-$n$ advantage arising from the smaller number of copies required by VAP is clearly observed. 
For the fixed subsystem geometries considered here, 
the LVP-to-VAP overhead ratio exceeds the copy-number factor $n$ 
at lower temperatures, indicating an additional reduction in sampling overhead. 
This comparison isolates the variance-related cost and does not account for 
possible increases in $d_\mathrm{buf}$ and $d_\mathrm{sep}$ required to maintain 
a fixed approximation bias as the correlation length grows. 

\begin{figure}
    \centering
    \includegraphics[width=\hsize]{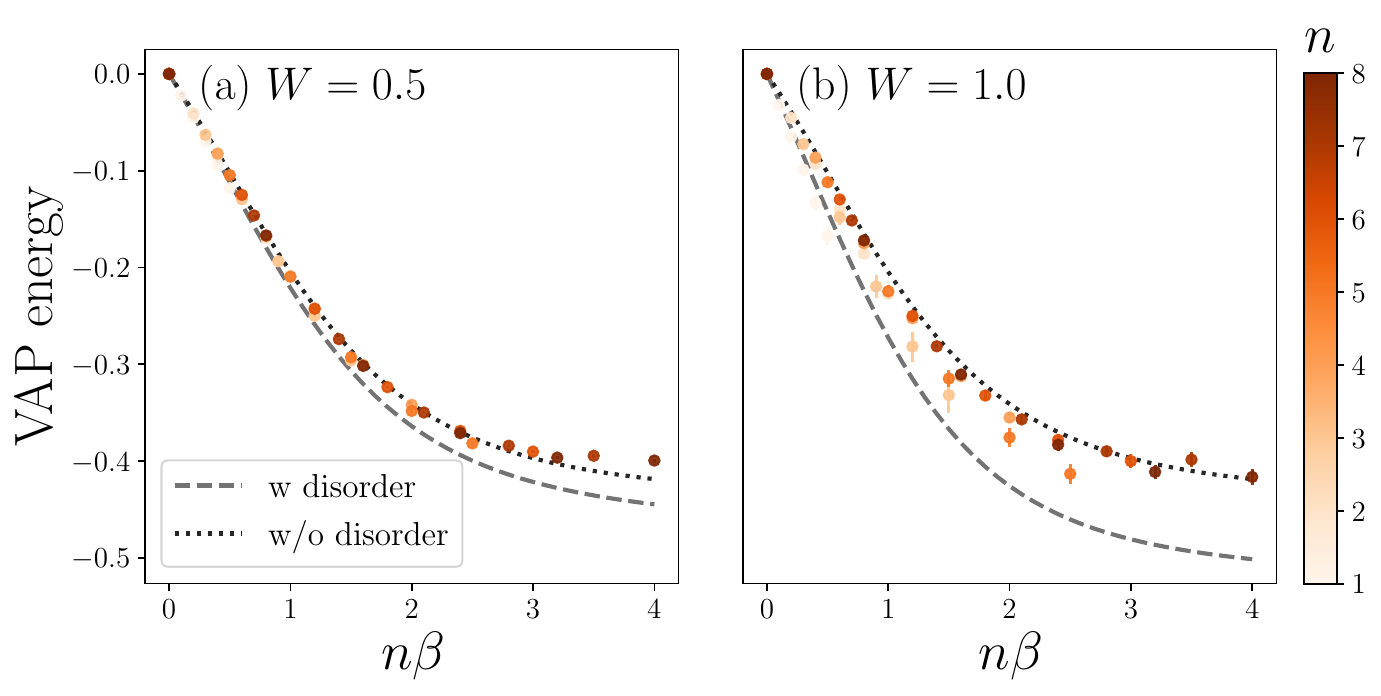}
    \caption{
    VAP energy for the random-field Heisenberg model for
    (a) $W=0.5$ and (b) $W=1.0$. 
    The data are averaged over 10 disorder realizations. 
    The gray dashed line indicates the energy of the original disordered thermal state, 
    while the black dotted line indicates the corresponding clean reference 
    without disorder $(N=50)$.
    }
    \label{fig:random}
\end{figure}

Finally, we numerically examine the robustness of VAP 
to violations of translational invariance in the Gibbs state 
due to miscalibration or algorithmic errors. 
To this end, we consider the one-dimensional Heisenberg model 
with a random magnetic field, 
\begin{align}
    H = \sum_{i=1}^{N-1} \bm{S}_i \cdot \bm{S}_{i+1}
    + \sum_{i=1}^N h_i S_i^z,
\end{align}
where $S_i^\alpha = \sigma_i^\alpha/2 \, (\alpha=x,y,z)$ denotes the spin operator at site $i$. 
Each $h_i$ is independently drawn from a uniform distribution on $[-W, W]$, 
where $W$ sets the strength of the random field. 
The random fields represent spatially varying calibration errors that cause 
the prepared state to be the Gibbs state of a Hamiltonian different from 
the target one. 
This setting allows us to test whether VAP suppresses the resulting 
deviation from the target Gibbs state.
Figure~\ref{fig:random} shows the numerical results for $W=0.5$ and $W=1.0$. 
In both cases, VAP substantially reduces the disorder-induced deviation of the energy, 
yielding values closer to the clean reference than to the disordered reference.
This behavior suggests that VAP has an effective disorder-averaging effect 
and can extract the approximately translationally invariant component of the thermal state.
A possible interpretation is 
that combining the density operators associated with multiple virtual copies 
acts more like averaging the underlying local Hamiltonians 
than directly averaging the density operators themselves.
Unlike FVP and LVP, which purify copies of the same spatially inhomogeneous 
state, VAP combines distinct spatial regions and can partially average out 
their local deviations. These results therefore suggest that VAP may mitigate 
spatially inhomogeneous errors that are not directly addressed by conventional 
virtual-purification protocols. 

\emph{\textbf{Summary and discussion.---}}
We have presented \emph{virtual autopurification} (VAP),
a protocol that performs virtual purification entirely
within a single translationally invariant quantum state.
By using spatially separated subregions as approximate virtual replicas,
VAP removes the need for multiple physical copies
while retaining the purification mechanism of fully virtual purification (FVP)
\cite{Cotler2019, Koczor2021, Huggins2021}.
Under clustering assumptions,
we showed that the VAP estimator reproduces the results of FVP and
localized virtual purification (LVP) \cite{Hakoshima2024},
with errors that decay exponentially with the buffer length
and the inter-copy separation.
Numerical simulations corroborated these predictions
and demonstrated accurate virtual cooling with reduced space--time overhead. 
Numerical results in the End Matter further show that VAP can simultaneously 
realize virtual cooling and mitigate local depolarizing noise, suggesting 
robustness against this form of decoherence noise. 

To implement virtual purification protocols on real quantum devices \cite{OBrien2023},
it is crucial to estimate nonlinear quantities
such as $\Tr[\rho^n]$ and $\Tr[\rho^n o_A]$ efficiently.
Beyond the naive generalized swap trick,
several approaches have recently been developed for this purpose,
including methods based on query access to a unitary 
that prepares a purification of $\rho$ \cite{Zhang2026},
randomized measurements \cite{Du2026, Li2026},
and classical shadows \cite{Seif2023, Liu2026}.
However, such estimator-level approaches do not by themselves remove
the exponential sampling overhead intrinsic to virtual purification.
In contrast, LVP and VAP exploit physical properties of $\rho$,
such as clustering and translational invariance,
to reduce this overhead at the level of the purification protocol itself.
This protocol-level reduction is complementary to nonlinear-estimation techniques
and to other error-mitigation or error-correction layers,
including quantum subspace expansion \cite{Yoshioka2022, Yang2025}
and approaches that combine virtual purification with quantum error correction \cite{Kwon2026}.
Related single-state approaches to virtual purification have also been proposed,
such as applying the inverse of the state-preparation circuit \cite{Huo2022}
or approximating the SWAP operator 
in the eigenbasis of the target observable \cite{Karim2026}.
Compared with these approaches,
VAP provides a rigorous error analysis controlled by physical properties of the state. 

The idea that a single thermal state can contain substantial information
about the underlying Hamiltonian has appeared in distinct contexts.
In quantum information,
Hamiltonian learning from Gibbs states has been actively studied
\cite{Anshu2021, Haah2024, Bakshi2024}.
In statistical mechanics,
related questions have been discussed for reconstructing a Hamiltonian
from a single thermal eigenstate \cite{Garrison2018}.
These perspectives suggest that a single thermal state can encode more information
than is directly accessible through conventional expectation values.
From this viewpoint, VAP can be regarded as a concrete protocol
for extracting lower-temperature information encoded in a single Gibbs state.
Exploring what other physical information can be accessed
from thermal states is an interesting direction for future work.

\emph{\textbf{Acknowledgments.---}}
A.~I. was supported by the Center of Innovation
for Sustainable Quantum AI (JST Grant No.~JPMJPF2221)
and JSPS KAKENHI (Grant No.~25K17308).
H.~H. was supported by the MEXT Quantum Leap Flagship Program 
(MEXT Q-LEAP) (Grant No.~JPMXS0120319794),
JST PRESTO (Grant No.~JPMJPR25F7),
and JSPS KAKENHI (Grant No.~JP24K16979).
N.~Y. was supported by JST (Grant No.~JPMJPF2221),
JST CREST (Grant No.~JPMJCR23I4),
JST ASPIRE (Grant No.~JPMJAP2316),
JST ERATO (Grant No.~JPMJER2302),
JST Moonshot R\&D (Grant No.~JPMJMS256J),
IBM Quantum, Google Quantum AI,
and the Institute for AI and Beyond at the University of Tokyo.

\emph{\textbf{Data availability.---}}
The code used to generate the numerical results reported in this work
is publicly available on GitHub \cite{VAPcode}.

\nocite{apsrev42Control}
\bibliographystyle{apsrev4-2}
\bibliography{references}

\section*{End matter}
\subsection*{Analysis under stochastic noise}

\begin{figure*}
    \centering
    \includegraphics[width=\hsize]{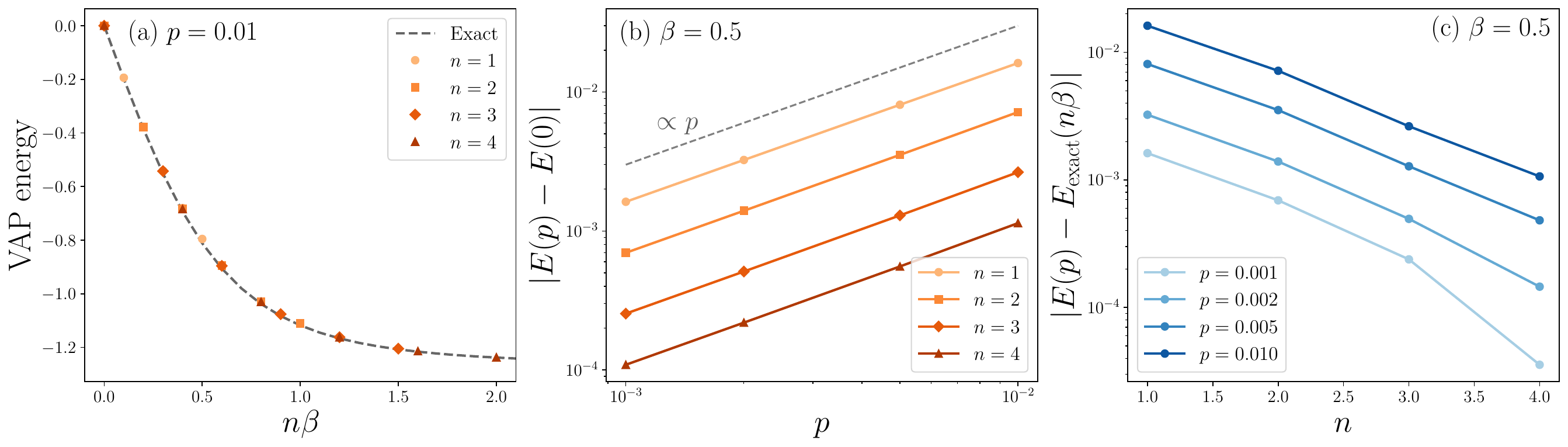}
    \caption{
    Thermal-state results in the presence of local depolarizing noise.
    (a) VAP energy rescaled by $n\beta$, plotted as a function of $n\beta$
    for $p=0.01$. The dashed line indicates the exact result.
    (b) Absolute deviation of the energy from its noiseless value at $p=0$,
    shown as a function of $p$ for $\beta=0.5$.
    (c) Absolute deviation of the energy from the exact result, shown as a
    function of $n$ for $\beta=0.5$.
    }
    \label{fig:noise}
\end{figure*}

We consider the effect of local depolarizing noise.
The depolarizing noise acting on the $i$th qubit is defined by
\begin{align}
    \mathcal{E}^{(i)}(\rho) = (1-p)\rho
    + \frac{p}{3}\left(X_i \rho X_i + Y_i \rho Y_i + Z_i \rho Z_i\right),
\end{align}
where $p$ denotes the error rate. The local depolarizing noise channel is
given by applying this noise independently to all sites:
\begin{align}
    \mathcal{E} = \bigotimes_{i=1}^N \mathcal{E}^{(i)}.
\end{align}
In the numerical calculations below, we consider a thermal state
$\rho_\beta \propto e^{-\beta H}$ subjected to local depolarizing noise,
and study a setting in which VAP simultaneously performs virtual cooling
and error mitigation.

We present numerical results for $d_\mathrm{buf}=4$ and $d_\mathrm{sep}=4$, with the
number of virtual copies varied from $n=1$ to $4$.
Figure~\ref{fig:noise}(a) shows the VAP energy plotted against $n\beta$
for a fixed error rate $p=0.01$. The dashed line denotes the exact
result, and the VAP data collapse well onto it.
Figure~\ref{fig:noise}(b) shows the deviation of the VAP energy from its
noiseless value at $p=0$. For all $n$, the noise-induced deviation grows
polynomially with $p$. Figure~\ref{fig:noise}(c) shows the deviation of
the VAP energy from the exact result. For all error rates considered,
this deviation is suppressed exponentially with increasing $n$.
Although we have not developed a theory of error mitigation in VAP, these
results numerically demonstrate that, under a realistic error model, VAP
can simultaneously realize virtual cooling and error mitigation.

To qualitatively understand Fig.~\ref{fig:noise}(b), 
consider a simple noninteracting toy model,
\begin{align}
    H = -\sum_i \sigma_i^z .
\end{align}
For this model, 
the thermal state factorizes into single-site mixed states with polarization 
$m = \langle \sigma^z \rangle =\tanh\beta$, 
and local depolarizing noise reduces the polarization as $m \to \eta m$ with $\eta = 1 - 4p/3$. 
Virtual purification with $n$ copies then gives an effective polarization
\begin{align}
    m_n = \tanh \!\left[n\, \mathrm{arctanh}(\eta m)\right],
\end{align}
so that, for small $p$, the energy deviation behaves as
\begin{align}
    \frac{|E(p)-E(0)|}{N} = C_n(\beta)\, p + \mathcal{O}(p^2) ,
\end{align}
where
\begin{align}
    C_n(\beta) = \frac{2n}{3}\sinh(2\beta)\,\mathrm{sech}^2(n\beta).
\end{align}
At finite $\beta$, 
the thermal state contains excited-state components 
whose weights are modified by stochastic noise. 
When these weights are raised to the $n$th power, 
a first-order noise correction in one factor combines 
with the noiseless thermal weight in the remaining factors.
As a result, 
a term linear in $p$ remains after virtual purification.
For a pure ground state, 
incoherent errors can introduce weight in orthogonal excited states. 
Virtual purification raises these noise-induced weights to the $n$th power, 
so their contribution to the error typically scales as $p^n$. 
In the toy model, 
the energy deviation in the pure ground-state limit $\beta\to\infty$ is
\begin{align}
    \frac{|E(p)-E(0)|}{N}
    &= 2\left(\frac{2p}{3}\right)^n + \mathcal{O}(p^{n+1}).
\end{align}
For $n\beta\gg1$, we have
$C_n(\beta)\sim(8n/3)\sinh(2\beta)e^{-2n\beta}$.
Thus, at fixed $\beta>0$, the coefficient decays exponentially
with $n$ up to the prefactor $n$, consistent with
Fig.~\ref{fig:noise}(c). 

\clearpage
\onecolumngrid
\setcounter{secnumdepth}{3}
\begin{center}
    \large
    \textbf{Supplemental Material for: \\ Virtual Autopurification in Translationally Invariant Quantum States}
\end{center}

\setcounter{section}{-1}
\setcounter{equation}{0}
\setcounter{figure}{0}
\setcounter{table}{0}
\renewcommand{\thesection}{S\arabic{section}}
\renewcommand{\theequation}{S\arabic{equation}}
\renewcommand{\thefigure}{S\arabic{figure}}
\renewcommand{\thetable}{S\arabic{table}}

\section{Setup and Notation}\label{sec:S0_setup}

\subsection{Setup}
We consider a quantum many-body system on a lattice equipped with the graph
distance $d$, with a local Hamiltonian
\begin{align}
    H = \sum_X h_X ,
\end{align}
where each term $h_X$ is supported on a subset $X$ of the lattice.
We assume that every nonzero interaction term satisfies $|X|\le k$ and
$\operatorname{diam}(X)=\max_{x,y\in X}d(x,y)\le r$.
We assume a uniform bound on the local interaction strength,
\begin{align}
    \max_{v} \sum_{X:\, v \in X} \|h_X\| \le g .
\end{align}
Here, $k$ bounds the number of sites involved in each interaction,
$r$ is the maximum spatial range of the interactions,
and $g$ sets the typical local energy scale.

\begin{figure*}[htbp]
    \centering
    \includegraphics[width=0.6\hsize]{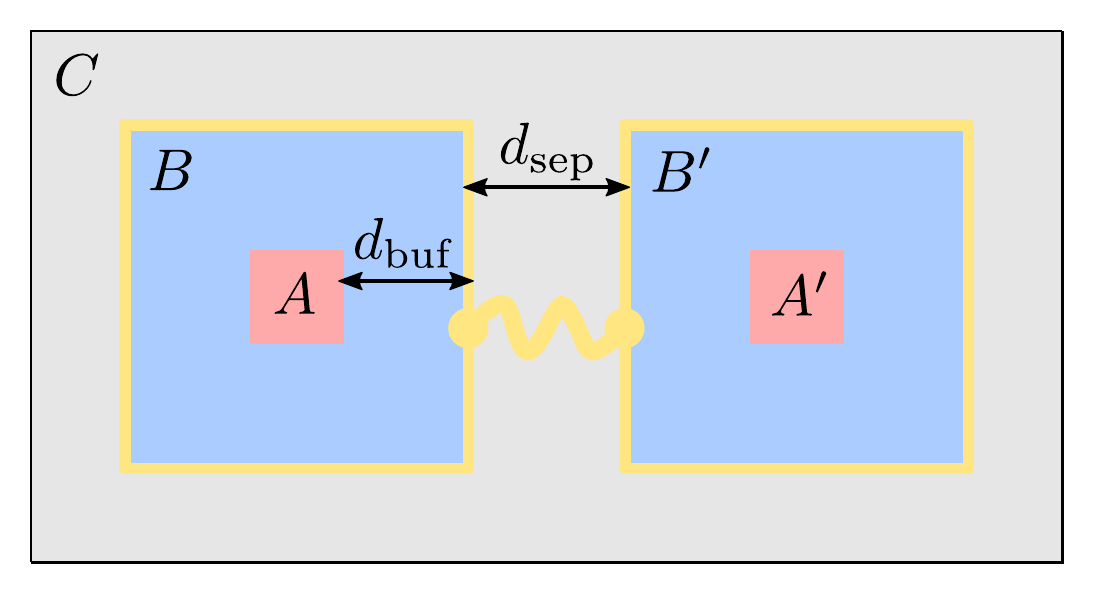}
    \caption{
    Schematic picture of the spatial partition used in the analysis. 
    Regions $B$ and $B'$ form buffers around $A$ and $A'$, respectively, 
    while $C$ denotes the remainder of the system. 
    The buffer length $d_\mathrm{buf}$ and the inter-copy separation 
    $d_\mathrm{sep}$ are also indicated.
    }
    \label{fig:S1_partition}
\end{figure*}

As illustrated schematically in Fig.~\ref{fig:S1_partition}, 
we partition the system into disjoint regions $A,B,A',B',C$.
Region $A$ supports the local observable of interest $o_A$.
We assume that its support contains at most $k$ sites:
\begin{equation}
    |\operatorname{supp}(o_A)| \le k .
    \label{eq:S0_observable_support}
\end{equation}
Region $B$ is a buffer surrounding $A$, and we set
\begin{align}
    AB = A \cup B .
\end{align}
We further choose $A'B'$ to be a translated copy of $AB$, spatially separated from it.
The remainder of the system is denoted by $C$.

Throughout the analysis, 
we use two characteristic distances: 
the buffer length $d_\mathrm{buf}$ between $A$ and $C$, 
and the inter-copy separation $d_\mathrm{sep}$ between $AB$ and $A'B'$.
These two length scales, 
shown in Fig.~\ref{fig:S1_partition}, 
govern the error of the LVP approximation 
and the additional error introduced in VAP, respectively. 
We assume $d_\mathrm{sep}>r$ 
to exclude direct interactions between $AB$ and $A'B'$.

We denote the mixed state of the full system by $\rho$.
For any region $R$, the reduced state on $R$ is written as
\begin{align}
    \rho_R = \Tr_{\bar{R}}[\rho] ,
\end{align}
where $\bar{R}$ is the complement of $R$.
When the global state is translationally invariant and $A'B'$ is a translated copy of $AB$, we have
\begin{align}
    \rho_{AB} = \rho_{A'B'} .
\end{align}

\subsection{Swap operator}
Let $S_{AB,A'B'}$ denote the swap operator exchanging the two isomorphic subsystems $AB$ and $A'B'$:
\begin{align}
    S_{AB,A'B'} \ket{\psi}_{AB}\otimes \ket{\phi}_{A'B'}
    = \ket{\phi}_{AB}\otimes \ket{\psi}_{A'B'} ,
\end{align}
with trivial action on the rest of the system.
We will repeatedly use the identity
\begin{align}
    \Tr_{ABA'B'}\!\bigl[(X_{AB}\otimes Y_{A'B'})\,S_{AB,A'B'}\bigr]
    = \Tr_{AB}[X_{AB}Y_{AB}] ,
    \label{eq:S0_swap_identity}
\end{align}
where $X_{AB}$ and $Y_{A'B'}$ act on isomorphic subsystems.

\subsection{LVP and VAP estimators (for $n=2$)}
For the remainder of the Supplemental Material, we focus on the case $n=2$.
Given a local observable $o_A$ supported on $A$, 
we define the localized virtual purification (LVP) expectation value as
\begin{align}
    \langle o_A\rangle_{\mathrm{LVP}}
    &= \frac{\Tr_{AB}[(\rho_{AB})^2\,o_A]}{\Tr_{AB}[(\rho_{AB})^2]} \notag\\
    &= \frac{\Tr_{ABA'B'}[\rho_{AB}\otimes \rho_{A'B'}\, S_{AB,A'B'}\, o_A]}
            {\Tr_{ABA'B'}[\rho_{AB}\otimes \rho_{A'B'}\, S_{AB,A'B'}]} .
    \label{eq:S0_LVP_def}
\end{align}
The virtual autopurification (VAP) estimator is defined by
\begin{align}
    \langle o_A\rangle_{\mathrm{VAP}}
    = \frac{\Tr_{ABA'B'}[\rho_{ABA'B'}\, S_{AB,A'B'}\, o_A]}
            {\Tr_{ABA'B'}[\rho_{ABA'B'}\, S_{AB,A'B'}]} .
    \label{eq:S0_VAP_def}
\end{align}
Using the identity in the previous subsection, one obtains the exact relation
\begin{align}
    \langle o_A\rangle_{\mathrm{VAP}}-\langle o_A\rangle_{\mathrm{LVP}}
    =
    \frac{\Tr_{ABA'B'}\!\left[(\rho_{ABA'B'}-\rho_{AB}\otimes \rho_{A'B'})\,
    S_{AB,A'B'}\,(o_A - \langle o_A\rangle_{\mathrm{VAP}})\right]}
    {\Tr_{AB}[(\rho_{AB})^2]} .
    \label{eq:S0_VAP_minus_LVP_identity}
\end{align}

\subsection{Perturbative setup}\label{sec:S0_perturbative_setup}
To control correlations between $AB$ and $A'B'$ through $C$, we decompose the
Hamiltonian into intra-region and boundary contributions:
\begin{align}
    H
    &= H_{AB}+H_{A'B'}+H_C+V_{BC}+V_{B'C} \notag\\
    &= H_0 + V ,
    \label{eq:S0_H_decomp}
\end{align}
Here, $H_0=H_{AB}+H_{A'B'}+H_C$ and $V=V_{BC}+V_{B'C}$.
For example, $H_{AB}=\sum_{X\subset AB} h_X$, and $V_{BC}$ collects interactions
straddling the boundary $\partial(BC)$:
\begin{equation}
    V_{BC}=\sum_{X\subset \partial(BC)} h_X ,
\end{equation}
and similarly for $V_{B'C}$.
We introduce a coupling parameter $\lambda$ as
\begin{equation}
    H(\lambda)=H_0+\lambda V,
\end{equation}
and define the corresponding Gibbs state
\begin{equation}
    \rho(\lambda)=\frac{e^{-\beta H(\lambda)}}{\Tr[e^{-\beta H(\lambda)}]} .
\end{equation}
The original Hamiltonian and Gibbs state are recovered as $H=H(\lambda = 1)$ and
$\rho=\rho(\lambda = 1)$. In Sec.~\ref{sec:S1_proof}, we expand the relevant quantities
around $\lambda=0$ and retain terms through second order in $\lambda$.
When no confusion can arise, we leave the $\lambda$ dependence
of reduced states implicit.

\section{Proof of Proposition~\ref{prop:VAP-LVP}}\label{sec:S1_proof}

In this section, we prove Proposition~\ref{prop:VAP-LVP} by a perturbative
expansion in the coupling $\lambda$ up to second order.
The main goal is to bound the difference between the VAP and LVP estimators
while keeping track of their normalization factors.
Throughout the proof, we use the perturbative setup introduced in
Sec.~\ref{sec:S0_setup}.

\subsection{Clustering assumption}

We begin by stating the clustering assumption used in the proof of
Proposition~\ref{prop:VAP-LVP}.
Throughout this section, we assume that the reduced density operators of
distant regions satisfy exponential clustering in trace norm:
\begin{align}
    \left\|\rho_{RR'}-\rho_R\otimes \rho_{R'}\right\|_1
    \le \mathrm{const.} \times e^{-d_{RR'}/\xi} ,
    \label{eq:S1_trace_norm_clustering}
\end{align}
for disjoint regions $R$ and $R'$, with some correlation length $\xi>0$.

This assumption immediately implies the standard exponential clustering of
connected correlations. Indeed, for operators $M_R$ and $M_{R'}$ supported on
$R$ and $R'$, respectively, we have
\begin{align}
    \left|\Tr[\rho\, M_R M_{R'}]-\Tr[\rho\, M_R]\Tr[\rho\, M_{R'}]\right|
    &=
    \left|\Tr\!\left[\left(\rho_{RR'}-\rho_R\otimes\rho_{R'}\right)
    (M_R\otimes M_{R'})\right]\right| \notag\\
    &\le
    \left\|\rho_{RR'}-\rho_R\otimes\rho_{R'}\right\|_1
    \|M_R\|\,\|M_{R'}\| \notag\\
    &\le
    \mathrm{const.} \times \|M_R\|\,\|M_{R'}\| ~ e^{-d_{RR'}/\xi} .
    \label{eq:S1_exp_clustering}
\end{align}

\subsection{Exact identity for the VAP-LVP difference}

We now rewrite the VAP-LVP difference in a form suitable for perturbative
analysis. For this purpose, we first establish explicitly that the first-order
contribution to the relevant density-operator difference vanishes. Let
$\rho_0=\rho(0)=e^{-\beta H_0}/\Tr[e^{-\beta H_0}]$, and denote its reduced
states on $AB$, $A'B'$, and $C$ by $\rho_{AB}^{(0)}$,
$\rho_{A'B'}^{(0)}$, and $\rho_C^{(0)}$, respectively. Since $H_0$ contains no
couplings between these regions, the unperturbed Gibbs state factorizes as
$\rho_0=\rho_{AB}^{(0)}\otimes\rho_{A'B'}^{(0)}\otimes\rho_C^{(0)}$.

The first-order correction to the normalized Gibbs state is linear in the
perturbation $V=V_{BC}+V_{B'C}$. Here and below, $\rho_{AB}^{(1)}$ and
$\rho_{A'B'}^{(1)}$ denote the coefficients of $\lambda$ in the expansions of
$\rho_{AB}(\lambda)$ and $\rho_{A'B'}(\lambda)$, respectively. If only $V_{BC}$
is retained, $A'B'$ remains decoupled from $ABC$ for all $\lambda$, so this
perturbation contributes only
$\rho_{AB}^{(1)}\otimes\rho_{A'B'}^{(0)}$ to the first-order correction of
$\rho_{ABA'B'}(\lambda)$. Similarly, $V_{B'C}$ contributes only
$\rho_{AB}^{(0)}\otimes\rho_{A'B'}^{(1)}$. Therefore,
\begin{align}
    \rho_{ABA'B'}(\lambda)
    &= \rho_{AB}^{(0)}\otimes\rho_{A'B'}^{(0)}
    +\lambda\left(
        \rho_{AB}^{(1)}\otimes\rho_{A'B'}^{(0)}
        +\rho_{AB}^{(0)}\otimes\rho_{A'B'}^{(1)}
    \right)
    +\mathcal{O}(\lambda^2),
    \label{eq:S1_joint_state_first_order}
\end{align}
On the other hand, expanding the product of the reduced states gives
\begin{align}
    \rho_{AB}(\lambda)\otimes\rho_{A'B'}(\lambda)
    &= \rho_{AB}^{(0)}\otimes\rho_{A'B'}^{(0)}
    +\lambda\left(
        \rho_{AB}^{(1)}\otimes\rho_{A'B'}^{(0)}
        +\rho_{AB}^{(0)}\otimes\rho_{A'B'}^{(1)}
    \right)
    +\mathcal{O}(\lambda^2).
    \label{eq:S1_product_state_first_order}
\end{align}
Thus, the zeroth- and first-order terms coincide, and hence
\begin{align}
    \rho_{ABA'B'}(\lambda)
    -\rho_{AB}(\lambda)\otimes\rho_{A'B'}(\lambda)
    =\mathcal{O}(\lambda^2).
    \label{eq:S1_first_order_cancellation}
\end{align}
Consequently, the difference
$\langle o_A\rangle_{\mathrm{VAP}}-\langle o_A\rangle_{\mathrm{LVP}}$
starts at second order in $\lambda$. Therefore, up to $\mathcal{O}(\lambda^2)$,
the normalization factors can be replaced by their zeroth-order values, which
yields
\begin{align}
    \langle o_A \rangle_\mathrm{VAP}
    - \langle o_A \rangle_\mathrm{LVP}
    &=
    \frac{\Tr_{ABA'B'}\!\left[(\rho_{ABA'B'} - \rho_{AB}\otimes\rho_{A'B'})\,
    S_{AB,A'B'}\,
    \left(o_A - \langle o_A \rangle_{\rho_0^2}\right)\right]}
    {\Tr_{AB}[(\rho_{AB}^{(0)})^2]}
    + \mathcal{O}(\lambda^3) .
    \label{eq:S1_diff_VAP_LVP}
\end{align}
Here
\begin{align}
    \langle o_A \rangle_{\rho_0^2}
    &= \frac{\Tr_{AB}[(\rho_{AB}^{(0)})^2\, o_A]}
           {\Tr_{AB}[(\rho_{AB}^{(0)})^2]} .
\end{align}

From this point on, we redefine $o_A$ by shifting it as
\begin{align}
    o_A \;\to\; o_A - \langle o_A \rangle_{\rho_0^2} .
\end{align}
With this convention,
\begin{align}
    \Tr_{AB}[(\rho_{AB}^{(0)})^2 o_A] = 0 .
\end{align}

\subsection{Second-order expansion}

Having established the cancellation of the first-order contribution, we now
evaluate the leading second-order term explicitly. We work in the interaction
picture with respect to $H_0$ and define
\begin{align}
    V^{\mathrm{I}}(s) = e^{s H_0} V e^{-s H_0}, \qquad s \in [0,\beta] .
\end{align}
For any operator $X$, we denote the thermal expectation value in the unperturbed
state $\rho_0$ by
\begin{align}
    \langle X \rangle_0 = \Tr[\rho_0 X] .
\end{align}
It is convenient to introduce the fluctuation operators
\begin{align}
    \delta V^{\mathrm{I}}(s) = V^{\mathrm{I}}(s) - \langle V \rangle_0 ,
\end{align}
and
\begin{align}
    \delta^2 V^{\mathrm{I}}(s)V^{\mathrm{I}}(t)
    = \delta V^{\mathrm{I}}(s)\delta V^{\mathrm{I}}(t)
    - \langle \delta V^{\mathrm{I}}(s)\delta V^{\mathrm{I}}(t)\rangle_0 .
\end{align}
With these definitions, the normalized Gibbs state admits the Duhamel expansion
\begin{align}
    \rho(\lambda)
    &=
    \rho_0
    - \lambda \rho_0 \int_0^\beta ds\, \delta V^{\mathrm{I}}(s)
    + \lambda^2 \rho_0 \int_0^\beta ds \int_0^s dt\,
    \delta^2 V^{\mathrm{I}}(s)V^{\mathrm{I}}(t)
    + \mathcal{O}(\lambda^3) .
    \label{eq:S1_rho_expansion}
\end{align}

We next trace out the region $C$ in order to isolate the effective coupling
between $AB$ and $A'B'$. In the subscripts of the effective operators, $A$
and $A'$ are omitted. The labels $B$ and $B'$ indicate the boundaries from
which these operators arise, although the operators may act on the full
regions $AB$ and $A'B'$, respectively. The effective boundary operators are
\begin{align}
    \tilde{V}_B(s)
    = \Tr_C[\rho_C^{(0)} V_{BC}^{\mathrm{I}}(s)], \qquad
    \tilde{V}_{B'}(s)
    = \Tr_C[\rho_C^{(0)} V_{B'C}^{\mathrm{I}}(s)],
\end{align}
where
\begin{align}
    \rho_C^{(0)} = \Tr_{ABA'B'}[\rho_0] .
\end{align}
We then define
\begin{align}
    \Delta_{BB'}(s,t)
    &=
    \Tr_C[\rho_C^{(0)} V_{BC}^{\mathrm{I}}(s)V_{B'C}^{\mathrm{I}}(t)]
    - \tilde{V}_B(s)\tilde{V}_{B'}(t) \notag\\
    &\quad
    + \Tr_C[\rho_C^{(0)} V_{B'C}^{\mathrm{I}}(s)V_{BC}^{\mathrm{I}}(t)]
    - \tilde{V}_{B'}(s)\tilde{V}_{B}(t) ,
\end{align}
which captures the connected second-order contribution between the two boundary
couplings through $C$.
From $\Delta_{BB'}(s,t)$, we further define
\begin{align}
    \Delta_B(s,t)
    &= \Tr_{A'B'}[\rho_{A'B'}^{(0)} \Delta_{BB'}(s,t)] , \\
    \Delta_{B'}(s,t)
    &= \Tr_{AB}[\rho_{AB}^{(0)} \Delta_{BB'}(s,t)] , \\
    \Delta(s,t)
    &= \langle \Delta_{BB'}(s,t)\rangle_0 .
\end{align}

Collecting all second-order terms, we obtain
\begin{align}
    \rho_{ABA'B'}-\rho_{AB}\otimes\rho_{A'B'}
    &=
    \lambda^2 \rho_{AB}^{(0)}\otimes\rho_{A'B'}^{(0)}
    \int_0^\beta ds \int_0^s dt\,
    \Bigl[
        \Delta_{BB'}(s,t)
        - \Delta_B(s,t)
        - \Delta_{B'}(s,t)
        + \Delta(s,t)
    \Bigr]
    + \mathcal{O}(\lambda^3) .
    \label{eq:S1_rho_diff}
\end{align}
Equation~\eqref{eq:S1_rho_diff} reduces the problem to bounding the four
second-order contributions separately, which we do in the next subsection.

\subsection{Bounds on the second-order terms}

\subsubsection{Reduction of the numerator}
Substituting Eq.~\eqref{eq:S1_rho_diff} into Eq.~\eqref{eq:S1_diff_VAP_LVP}, we find that the estimator difference is determined by four second-order contributions associated with $\Delta_{BB'}(s,t)$, $\Delta_B(s,t)$, $\Delta_{B'}(s,t)$, and $\Delta(s,t)$.
To evaluate these terms, it is convenient to introduce the imaginary-time evolved operator
\begin{align}
    O(\tau)=e^{-\tau H_{AB}} O e^{\tau H_{AB}} .
\end{align}
In particular, we will use $o_A(\beta)=e^{-\beta H_{AB}}o_A e^{\beta H_{AB}}$.

It is also convenient to rewrite the relevant traces in terms of the auxiliary state
\begin{align}
    \tilde{\rho}=
    \frac{\exp(-2\beta H_{AB}-\beta H_{A'B'}-\beta H_C)}
    {\Tr[\exp(-2\beta H_{AB}-\beta H_{A'B'}-\beta H_C)]} ,
\end{align}
which naturally appears after applying the swap identity and collecting the thermal weights.
We denote expectation values with respect to $\tilde{\rho}$ by $\langle\cdots\rangle_{\tilde{\rho}}$.

\subsubsection{Vanishing of the scalar term $\Delta(s,t)$}
We first note that the contribution proportional to $\Delta(s,t)$ vanishes.
Indeed, since $\Delta(s,t)$ is a scalar, it can be pulled out of the trace, and we obtain
\begin{align}
    \Tr[\rho_{AB}^{(0)} \otimes \rho_{A'B'}^{(0)}\, \Delta(s,t)\, S_{AB,A'B'}\, o_A]
    = \Delta(s,t)\, \Tr[(\rho_{AB}^{(0)})^2 o_A] .
\end{align}
By construction, the observable $o_A$ has been centered so that
\begin{align}
    \Tr[(\rho_{AB}^{(0)})^2 o_A] = 0 .
\end{align}
Therefore, the scalar contribution $\Delta(s,t)$ does not contribute to the
VAP--LVP difference at order $\lambda^2$.

\subsubsection{Bounds on the single-boundary terms $\Delta_B(s,t)$ and $\Delta_{B'}(s,t)$}
We first consider the contribution containing $\Delta_B(s,t)$.
Applying the swap identity in Eq.~\eqref{eq:S0_swap_identity} to the
normalized trace in Eq.~\eqref{eq:S1_diff_VAP_LVP}, and identifying the
unperturbed Gibbs states of the translated copies $AB$ and $A'B'$ under
the swap, we obtain
\begin{align*}
    &\frac{
    \Tr_{ABA'B'}[\rho_{AB}^{(0)} \otimes \rho_{A'B'}^{(0)}\,
    \Delta_B(s,t)\, S_{AB,A'B'}\, o_A]
    }{\Tr_{AB}[(\rho_{AB}^{(0)})^2]} \\
    &= \frac{\Tr_{AB}[\rho_{AB}^{(0)}\Delta_B(s,t)
    \rho_{AB}^{(0)}o_A]}{\Tr_{AB}[(\rho_{AB}^{(0)})^2]} \\
    &= \frac{\Tr_{AB}[(\rho_{AB}^{(0)})^2
    \Delta_B(s,t)o_A(\beta)]}{\Tr_{AB}[(\rho_{AB}^{(0)})^2]} \\
    &= \langle \Delta_{BB'}(s,t)o_A(\beta)\rangle_{\tilde{\rho}} .
\end{align*}
In the second equality, we used
$\rho_{AB}^{(0)}o_A=o_A(\beta)\rho_{AB}^{(0)}$ and the cyclicity of the trace.
The last equality follows from the definition of $\Delta_B(s,t)$ as a trace
over $A'B'$ and the definition of $\tilde{\rho}$.
Expanding $\Delta_{BB'}(s,t)$, we obtain
\begin{align}
    &\langle \Delta_{BB'}(s,t)o_A(\beta)\rangle_{\tilde{\rho}} \notag\\
    &=
    \langle V_{BC}^{\mathrm{I}}(s)\, o_A(\beta)\, V_{B'C}^{\mathrm{I}}(t) \rangle_{\tilde{\rho}}
    - \langle V_{BC}^{\mathrm{I}}(s)\, o_A(\beta) \rangle_{\tilde{\rho}}
      \langle V_{B'C}^{\mathrm{I}}(t) \rangle_{\tilde{\rho}} \notag\\
    &\quad
    + \langle V_{B'C}^{\mathrm{I}}(s)\, V_{BC}^{\mathrm{I}}(t)\, o_A(\beta) \rangle_{\tilde{\rho}}
    - \langle V_{B'C}^{\mathrm{I}}(s) \rangle_{\tilde{\rho}}
      \langle V_{BC}^{\mathrm{I}}(t)\, o_A(\beta) \rangle_{\tilde{\rho}} .
    \label{eq:S1_single_boundary_rewrite}
\end{align}
Here, the products of expectation values arise from the factorization of
$\tilde{\rho}$ between $AB$, $A'B'$, and $C$. In the first term, we used
$[o_A(\beta),V_{B'C}^{\mathrm{I}}(t)]=0$, since these operators act on
disjoint regions.
The contribution containing $\Delta_{B'}(s,t)$ is treated in the same way.

To bound the connected correlators in Eq.~\eqref{eq:S1_single_boundary_rewrite},
we use a quasi-locality property of the imaginary-time evolved boundary
interactions. Roughly speaking, although $V_{BC}^{\mathrm{I}}(s)$ and
$V_{B'C}^{\mathrm{I}}(t)$ are not
strictly local, they can be approximated by operators supported near the
corresponding boundaries, with errors decaying exponentially in the truncation
distance. This allows us to combine quasi-locality with exponential clustering
to obtain an exponential bound in the inter-copy distance $d_\mathrm{sep}$.

The following lemma is closely related to standard quasi-locality estimates for
imaginary-time evolution. For completeness, we provide a self-contained proof
below. The argument follows the usual nested-commutator expansion together with
standard bounds on multicommutators; see, e.g., Appendix E.1 and Lemma 13 of
Ref.~\cite{Kuwahara2021}.

\begin{lem}\label{lem:ht_quasilocal_hX}
Let $h_X$ be a local Hamiltonian term supported on $X$, and let $R$ be a
region containing $X$. Define
\begin{align}
    h_X^{(R)}(\beta) = e^{-\beta H_R} \, h_X \, e^{\beta H_R} ,
\end{align}
where $H_R$ denotes the restriction of the Hamiltonian to the region $R$.
Assume $2gk\beta < 1$, and let
\begin{align}
    D = \left\lceil \frac{d(X,\bar{R})}{r} \right\rceil , 
\end{align}
where $\bar{R}$ is the complement of $R$. 
Then
\begin{align}
    \| h_X(\beta) - h_X^{(R)}(\beta) \|
    \le
    2g \, \frac{(2gk\beta)^D}{1-2gk\beta} ,
    \label{eq:S1_ht_quasilocal_hX}
\end{align}
where
\begin{align}
    h_X(\beta) = e^{-\beta H} \, h_X \, e^{\beta H} .
\end{align}
\end{lem}

\begin{proof}
We compare the full imaginary-time evolution generated by $H$ with the truncated
one generated by $H_R$. Expanding both in nested commutators, we obtain
\begin{align}
    h_X(\beta) - h_X^{(R)}(\beta)
    &=
    \sum_{n=0}^{\infty} \frac{(-\beta)^n}{n!}
    \left(
        \mathrm{ad}_H^{(n)}[h_X]
        -
        \mathrm{ad}_{H_R}^{(n)}[h_X]
    \right) ,
    \label{eq:S1_series_diff}
\end{align}
where $\mathrm{ad}_K(O)=[K,O]$ and
$\mathrm{ad}_K^{(n)}=\mathrm{ad}_K\circ\cdots\circ \mathrm{ad}_K$.

Each $n$-fold commutator in Eq.~\eqref{eq:S1_series_diff} can be written as a
sum over sequences of local Hamiltonian terms whose supports overlap
successively, starting from the original support $X$ of $h_X$.
Since each Hamiltonian term has spatial diameter at most $r$, the support of
the operator can expand by at most distance $r$ at each commutator step.

Therefore, if
\begin{align}
    n < D = \left\lceil \frac{d(X,\bar{R})}{r} \right\rceil ,
\end{align}
no sequence of $n$ overlapping local terms can reach the complement $\bar{R}$.
In that case, every term contributing to $\mathrm{ad}_H^{(n)}[h_X]$ is already
contained entirely inside $R$, and hence coincides with the corresponding term
in $\mathrm{ad}_{H_R}^{(n)}[h_X]$. It follows that
\begin{align}
    \mathrm{ad}_H^{(n)}[h_X]
    =
    \mathrm{ad}_{H_R}^{(n)}[h_X]
    \qquad
    (n<D) .
\end{align}
Thus the difference in Eq.~\eqref{eq:S1_series_diff} receives contributions
only from orders $n \ge D$.

For the remaining terms with $n \ge D$, we bound the norm of the difference
naively by estimating each nested commutator separately. Using
\begin{align}
    \|\mathrm{ad}_A(B)\| = \|[A,B]\| \le 2 \|A\| \, \|B\| ,
\end{align}
together with the assumption
\begin{align}
    \max_v \sum_{Y:\, v \in Y} \|h_Y\| \le g ,
\end{align}
and the fact that each interaction term acts on at most $k$ sites, one obtains
the standard estimate
\begin{align}
    \bigl\|\mathrm{ad}_H^{(n)}[h_X]\bigr\|
    \le 2g \, (2gk)^n \, n! ,
\end{align}
and similarly for $\mathrm{ad}_{H_R}^{(n)}[h_X]$. Hence
\begin{align}
    \bigl\|
    \mathrm{ad}_H^{(n)}[h_X]
    -
    \mathrm{ad}_{H_R}^{(n)}[h_X]
    \bigr\|
    \le
    2g \, (2gk)^n \, n! .
\end{align}
Substituting this into Eq.~\eqref{eq:S1_series_diff}, we obtain
\begin{align}
    \|h_X(\beta) - h_X^{(R)}(\beta)\|
    &\le
    \sum_{n=D}^{\infty}
    \frac{\beta^n}{n!}
    \bigl\|
    \mathrm{ad}_H^{(n)}[h_X]
    -
    \mathrm{ad}_{H_R}^{(n)}[h_X]
    \bigr\| \notag\\
    &\le
    2g \sum_{n=D}^{\infty} (2gk\beta)^n .
\end{align}
Since $2gk\beta < 1$, the sum is geometric and converges:
\begin{align}
    2g \sum_{n=D}^{\infty} (2gk\beta)^n
    =
    2g \, \frac{(2gk\beta)^D}{1-2gk\beta} .
\end{align}
This proves
\begin{align}
    \| h_X(\beta) - h_X^{(R)}(\beta) \|
    \le
    2g \, \frac{(2gk\beta)^D}{1-2gk\beta} .
\end{align}
\end{proof}

The same estimate holds for the opposite convention
$e^{\beta H}h_Xe^{-\beta H}$, since replacing $\beta$ by $-\beta$ changes
only the signs of the nested-commutator coefficients and not their norm bounds.

Using this opposite-sign counterpart of
Lemma~\ref{lem:ht_quasilocal_hX}, we truncate the imaginary-time evolved
boundary operators $V_{BC}^{\mathrm{I}}(s)$ and $V_{B'C}^{\mathrm{I}}(t)$
within $C$. Let $R_1,R_2\subset C$ denote the regions retained when
truncating at distances $\ell_1$ and $\ell_2$ from the $BC$ and $B'C$
boundaries, respectively. The corresponding truncated operators are
denoted by $V_{BC}^{\mathrm{I}(R_1)}(s)$ and $V_{B'C}^{\mathrm{I}(R_2)}(t)$.
Since
\begin{align}
    V_{BC}^{\mathrm{I}}(s)=\sum_{X\subset \partial(BC)} h_X^{\mathrm{I}}(s) ,
\end{align}
Lemma~\ref{lem:ht_quasilocal_hX} implies
\begin{align*}
    \| V_{BC}^{\mathrm{I}}(s) - V_{BC}^{\mathrm{I}(R_1)}(s) \|
    &\le \sum_{X\subset \partial(BC)}
    \| h_X^{\mathrm{I}}(s) - h_X^{\mathrm{I}(R_1)}(s) \| \\
    &\le 2g |\partial B| \frac{(2gks)^{\ell_1/r}}{1-2gks} .
\end{align*}
Therefore,
\begin{equation}
    \| V_{BC}^{\mathrm{I}}(s) - V_{BC}^{\mathrm{I}(R_1)}(s) \|
    \le \mathrm{const.} \times |\partial B| ~ e^{-\ell_1 / \xi_s} ,
\end{equation}
where
\begin{equation}
    \xi_s = \frac{r}{\log(1/2gks)} .
\end{equation}
Similarly,
\begin{align}
    \| V_{B'C}^{\mathrm{I}}(t) - V_{B'C}^{\mathrm{I}(R_2)}(t) \|
    \le \mathrm{const.} \times |\partial B'| ~ e^{-\ell_2 / \xi_t} .
    \label{eq:quasi_local}
\end{align}

Moreover, using the support condition in Eq.~\eqref{eq:S0_observable_support}
and Eq.~(28) of Ref.~\cite{Alhambra2023}, for $2gk\beta<1$ we have
\begin{equation}
    \|o_A(\beta)\| \le k \| o_A \| \frac{1}{1-2gk\beta}
    = \mathrm{const.} \times \|o_A\| .
\end{equation}

Combining these bounds yields
\begin{align}
    &|\langle V_{BC}^{\mathrm{I}}(s) \, o_A(\beta) \,
    V_{B'C}^{\mathrm{I}}(t) \rangle_{\tilde{\rho}}
    - \langle V_{BC}^{\mathrm{I}}(s) \, o_A(\beta) \rangle_{\tilde{\rho}} \,
    \langle V_{B'C}^{\mathrm{I}}(t) \rangle_{\tilde{\rho}}| \notag \\
    &\le
    \mathrm{const.} \times |\partial B| \, \|o_A\| \, \|V_{B'C}\| \, e^{-\ell_1 / \xi_s} \notag \\
    &\quad
    + \mathrm{const.} \times |\partial B'| \, \|o_A\| \, \|V_{BC}\| \, e^{-\ell_2 / \xi_t} \notag \\
    &\quad
    + |\langle
    V_{BC}^{\mathrm{I}(R_1)}(s) \, o_A(\beta) \, V_{B'C}^{\mathrm{I}(R_2)}(t) \rangle_{\tilde{\rho}}
    - \langle V_{BC}^{\mathrm{I}(R_1)}(s) \, o_A(\beta) \rangle_{\tilde{\rho}} \,
    \langle V_{B'C}^{\mathrm{I}(R_2)}(t) \rangle_{\tilde{\rho}}| .
\end{align}

Although $o_A(\beta)$ need not be supported on $A$, it remains supported
within $AB$, because its imaginary-time evolution is generated by $H_{AB}$ alone.
At this stage, the weaker clustering assumption is sufficient.
Using the exponential clustering of connected correlations introduced in
Eq.~\eqref{eq:S1_exp_clustering}, we obtain
\begin{align}
    &|\langle V_{BC}^{\mathrm{I}(R_1)}(s) \, o_A(\beta) \, V_{B'C}^{\mathrm{I}(R_2)}(t) \rangle_{\tilde{\rho}}
    - \langle V_{BC}^{\mathrm{I}(R_1)}(s) \, o_A(\beta) \rangle_{\tilde{\rho}} \,
    \langle V_{B'C}^{\mathrm{I}(R_2)}(t) \rangle_{\tilde{\rho}}| \notag \\
    &\le
    \mathrm{const.} \times \|o_A(\beta)\| \, \|V_{BC}^{\mathrm{I}(R_1)}(s)\| \, \|V_{B'C}^{\mathrm{I}(R_2)}(t)\| \,
    e^{-(d_\mathrm{sep}-\ell_1-\ell_2)/\tilde{\xi}} \notag \\
    &\le
    \mathrm{const.} \times \|o_A\| \, \|V_{BC}\| \, \|V_{B'C}\| \,
    e^{-(d_\mathrm{sep}-\ell_1-\ell_2)/\tilde{\xi}} .
\end{align}
Substituting this into the previous bound, we obtain
\begin{align}
    &|\langle V_{BC}^{\mathrm{I}}(s) \, o_A(\beta) \,
    V_{B'C}^{\mathrm{I}}(t) \rangle_{\tilde{\rho}}
    - \langle V_{BC}^{\mathrm{I}}(s) \, o_A(\beta) \rangle_{\tilde{\rho}} \,
    \langle V_{B'C}^{\mathrm{I}}(t) \rangle_{\tilde{\rho}}| \notag \\
    &\le
    \mathrm{const.} \times |\partial B| \, |\partial B'| \, \|o_A\| \,
    (e^{-\ell_1 / \xi_s} + e^{-\ell_2 / \xi_t}) \notag \\
    &\quad
    + \mathrm{const.} \times \|o_A\| \, \|V_{BC}\| \, \|V_{B'C}\| \,
    e^{-(d_\mathrm{sep}-\ell_1-\ell_2)/\tilde{\xi}} .
\end{align}
Choosing
\begin{align}
    \ell_1 = \frac{\xi_s}{\tilde{\xi} + \xi_s + \xi_t} \, d_\mathrm{sep} , \qquad
    \ell_2 = \frac{\xi_t}{\tilde{\xi} + \xi_s + \xi_t} \, d_\mathrm{sep} ,
\end{align}
and defining
\begin{align}
    \xi' = \tilde{\xi} + \xi_s + \xi_t ,
\end{align}
we arrive at
\begin{align}
    &|\langle V_{BC}^{\mathrm{I}}(s) \, o_A(\beta) \,
    V_{B'C}^{\mathrm{I}}(t) \rangle_{\tilde{\rho}}
    - \langle V_{BC}^{\mathrm{I}}(s) \, o_A(\beta) \rangle_{\tilde{\rho}} \,
    \langle V_{B'C}^{\mathrm{I}}(t) \rangle_{\tilde{\rho}}| \notag \\
    &\le
    \mathrm{const.} \times |\partial B| \, |\partial B'| \, \|o_A\| \, e^{-d_\mathrm{sep} / \xi'} .
\end{align}
Applying the same reasoning to the other connected term in
Eq.~\eqref{eq:S1_single_boundary_rewrite}, we conclude that the contributions
containing $\Delta_B(s,t)$ and $\Delta_{B'}(s,t)$ are exponentially suppressed
in $d_\mathrm{sep}$.

\subsubsection{Bound on the double-boundary term $\Delta_{BB'}(s,t)$}
We finally consider the contribution containing $\Delta_{BB'}(s,t)$.
In contrast to the single-boundary terms, the estimate of this term requires
the stronger trace-norm clustering assumption introduced in
Eq.~\eqref{eq:S1_trace_norm_clustering}.

\begin{figure}
    \centering
    \includegraphics[width=0.5\hsize]{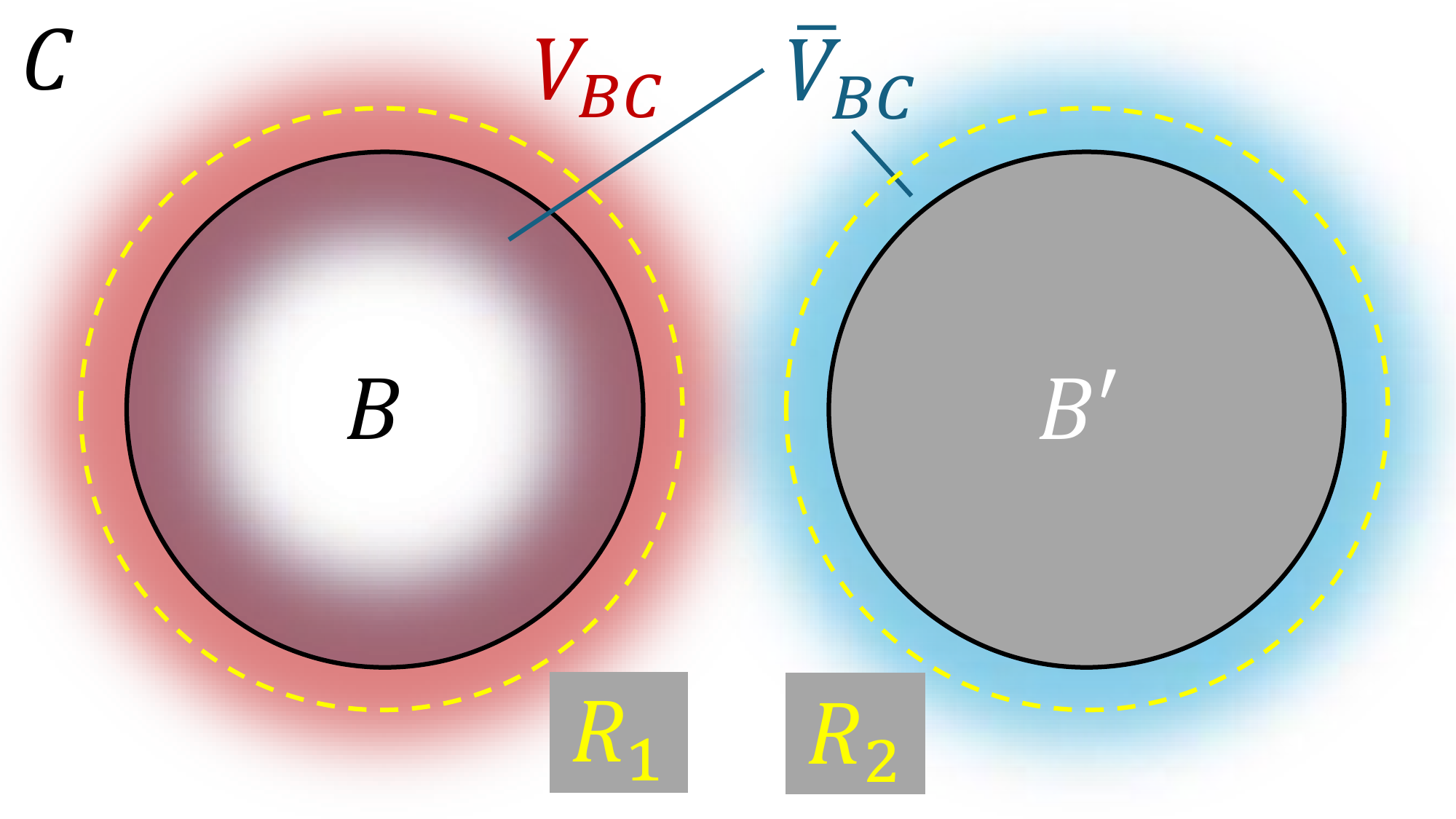}
    \caption{
    Schematic illustration of the supports of $V_{BC}$ and $\bar{V}_{BC}$.
    The operator $V_{BC}$ acts on the boundary region between $B$ and $C$, shown in light red.
    The operator $\bar{V}_{BC}$ acts on the interior of $B$ and on the part of $C$ adjacent to $B'$, shown in light blue.
    The regions $R_1$ and $R_2$ indicate the parts of $C$ retained by the corresponding truncations.
    }
    \label{fig:vbc_region}
\end{figure}

\begin{samepage}
Including the normalization factor in
Eq.~\eqref{eq:S1_diff_VAP_LVP}, we define
\begin{align}
    \tilde{\rho}_{AB}
    = \frac{(\rho_{AB}^{(0)})^2}{\Tr_{AB}[(\rho_{AB}^{(0)})^2]} ,
\end{align}
and $\bar{V}_{BC}(s, t)$ such that it satisfies
\begin{align}
    I_A \otimes \bar{V}_{BC} \otimes I_{A'B'} 
    = S_{AB,A'B'} \, ( I_{ABA'} \otimes V_{B'C}) \, S_{AB,A'B'} ,
    \qquad
    \bar{V}_{BC}(s,t)
    = e^{-sH_{AB}-tH_C} \, \bar{V}_{BC} \, e^{sH_{AB}+tH_C} .
\end{align}
\end{samepage}

The swap identity in Eq.~\eqref{eq:S0_swap_identity} places the
$AB$ factor of $V_{BC}^{\mathrm{I}}$ first in both cross terms.
Moving a Gibbs factor through the swapped interaction and using
$\rho_{AB}^{(0)}o_A=o_A(\beta)\rho_{AB}^{(0)}$ gives, for the first
ordering,
\begin{align*}
    S_{AB,A'B'}V_{B'C}^{\mathrm{I}}(t)S_{AB,A'B'}
    &= \bar{V}_{BC}(-t,-t),\\
    \rho_{AB}^{(0)}\bar{V}_{BC}(-t,-t)
    (\rho_{AB}^{(0)})^{-1}
    &= \bar{V}_{BC}(\beta-t,-t).
\end{align*}
For the opposite ordering, the $C$ factors must also be reordered
inside their thermal trace. Thermal cyclicity shifts their time
argument, so that
\begin{align*}
    \rho_C^{(0)}\bar{V}_{BC}(-s,-s)
    &= \bar{V}_{BC}(-s,\beta-s)\rho_C^{(0)},\\
    \rho_{AB}^{(0)}\bar{V}_{BC}(-s,\beta-s)
    (\rho_{AB}^{(0)})^{-1}
    &= \bar{V}_{BC}(\beta-s,\beta-s).
\end{align*}
The factorizable subtraction terms transform in the same way.
Combining the two orderings, we obtain
\begin{align}
    &\frac{
    \Tr_{ABA'B'}[\rho_{AB}^{(0)} \otimes \rho_{A'B'}^{(0)} \, \Delta_{BB'}(s,t)
    \, S_{AB,A'B'} \, o_A]
    }{\Tr_{AB}[(\rho_{AB}^{(0)})^2]} \notag\\
    &= \Tr_{AB}[\tilde{\rho}_{AB}
    \{ \Tr_C [\rho_C^{(0)} \, V_{BC}^{\mathrm{I}}(s) \, \bar{V}_{BC}(\beta-t,-t)]
    - \Tr_C[\rho_C^{(0)} \, V_{BC}^{\mathrm{I}}(s)] \,
      \Tr_C[\rho_C^{(0)} \, \bar{V}_{BC}(\beta-t,-t)] \}
    \, o_A(\beta)] \notag\\
    &\quad
    + \Tr_{AB}[\tilde{\rho}_{AB}
    \{ \Tr_C [\rho_C^{(0)} \, V_{BC}^{\mathrm{I}}(t) \, \bar{V}_{BC}(\beta-s,\beta-s)]
    - \Tr_C[\rho_C^{(0)} \, V_{BC}^{\mathrm{I}}(t)] \,
      \Tr_C[\rho_C^{(0)} \, \bar{V}_{BC}(\beta-s,\beta-s)] \}
    \, o_A(\beta)] ,
    \label{eq:S1_double_boundary_rewrite}
\end{align}
Figure~\ref{fig:vbc_region} schematically shows the supports of $V_{BC}$ and $\bar{V}_{BC}$.
In particular, note that the two operators have overlapping support in region $B$.

For the first term in Eq.~\eqref{eq:S1_double_boundary_rewrite}, we bound
\begin{align}
    &\Bigl| \Tr_{AB}[\tilde{\rho}_{AB}
    \{ \Tr_C [\rho_C^{(0)} \, V_{BC}^{\mathrm{I}}(s) \, \bar{V}_{BC}(\beta-t,-t)]
    - \Tr_C[\rho_C^{(0)} \, V_{BC}^{\mathrm{I}}(s)] \,
      \Tr_C[\rho_C^{(0)} \, \bar{V}_{BC}(\beta-t,-t)] \}
    \, o_A(\beta)] \Bigr| \notag\\
    &\le
    \mathrm{const.} \times \|o_A\| \,
    \| \Tr_C [\rho_C^{(0)} \, V_{BC}^{\mathrm{I}}(s) \, \bar{V}_{BC}(\beta-t,-t)]
    - \Tr_C[\rho_C^{(0)} \, V_{BC}^{\mathrm{I}}(s)] \,
      \Tr_C[\rho_C^{(0)} \, \bar{V}_{BC}(\beta-t,-t)] \| .
\end{align}
Applying the same quasi-local truncation as above, we obtain
\begin{align}
    &\| \Tr_C [\rho_C^{(0)} \, V_{BC}^{\mathrm{I}}(s) \, \bar{V}_{BC}(\beta-t,-t)]
    - \Tr_C[\rho_C^{(0)} \, V_{BC}^{\mathrm{I}}(s)] \,
      \Tr_C[\rho_C^{(0)} \, \bar{V}_{BC}(\beta-t,-t)] \| \notag\\
    &\le
    \mathrm{const.} \times |\partial B| \, |\partial B'|
    \left( e^{-\ell_1/\xi_s} + e^{-\ell_2/\xi_t} \right) \notag\\
    &\quad
    + \| \Tr_C [\rho_C^{(0)} \, V_{BC}^{\mathrm{I}(R_1)}(s) \,
    \bar{V}_{BC}^{(R_2)}(\beta-t,-t)]
    - \Tr_C[\rho_C^{(0)} \, V_{BC}^{\mathrm{I}(R_1)}(s)] \,
      \Tr_C[\rho_C^{(0)} \, \bar{V}_{BC}^{(R_2)}(\beta-t,-t)] \| .
\end{align}

For the last term, 
the stronger trace-norm clustering assumption in Eq.\eqref{eq:clustering_state} is needed.
Tracing over the truncation regions $R_1$ and $R_2$ shown in
Fig.~\ref{fig:vbc_region}, we have
\begin{align}
    &\| \Tr_C [\rho_C^{(0)} \, V_{BC}^{\mathrm{I}(R_1)}(s) \,
    \bar{V}_{BC}^{(R_2)}(\beta-t,-t)]
    - \Tr_C[\rho_C^{(0)} \, V_{BC}^{\mathrm{I}(R_1)}(s)] \,
      \Tr_C[\rho_C^{(0)} \, \bar{V}_{BC}^{(R_2)}(\beta-t,-t)] \| \notag\\
    &= \| \Tr_{R_1 R_2}
    [(\rho_{R_1 R_2}^{(0)} - \rho_{R_1}^{(0)} \otimes \rho_{R_2}^{(0)}) \,
    V_{BC}^{\mathrm{I}(R_1)}(s) \, \bar{V}_{BC}^{(R_2)}(\beta-t,-t)] \| .
\end{align}
Using Eq.~\eqref{eq:S1_trace_norm_clustering}, this is bounded by
\begin{align}
    &\| \Tr_{R_1 R_2}
    [(\rho_{R_1 R_2}^{(0)} - \rho_{R_1}^{(0)} \otimes \rho_{R_2}^{(0)}) \,
    V_{BC}^{\mathrm{I}(R_1)}(s) \, \bar{V}_{BC}^{(R_2)}(\beta-t,-t)] \| \notag\\
    &\le
    \mathrm{const.} \times
    \|V_{BC}^{\mathrm{I}(R_1)}(s)\| \,
    \|\bar{V}_{BC}^{(R_2)}(\beta-t,-t)\| \,
    e^{-(d_\mathrm{sep}-\ell_1-\ell_2)/\xi^*} \notag\\
    &\le
    \mathrm{const.} \times
    \|V_{BC}\| \, \|V_{B'C}\| \,
    e^{-(d_\mathrm{sep}-\ell_1-\ell_2)/\xi^*} .
\end{align}

Substituting this back into the previous estimate, we obtain
\begin{align}
    &\| \Tr_C [\rho_C^{(0)} \, V_{BC}^{\mathrm{I}}(s) \, \bar{V}_{BC}(\beta-t,-t)]
    - \Tr_C[\rho_C^{(0)} \, V_{BC}^{\mathrm{I}}(s)] \,
      \Tr_C[\rho_C^{(0)} \, \bar{V}_{BC}(\beta-t,-t)] \| \notag\\
    &\le
    \mathrm{const.} \times |\partial B| \, |\partial B'|
    \left( e^{-\ell_1/\xi_s} + e^{-\ell_2/\xi_t} \right) \notag\\
    &\quad
    + \mathrm{const.} \times
    \|V_{BC}\| \, \|V_{B'C}\| \,
    e^{-(d_\mathrm{sep}-\ell_1-\ell_2)/\xi^*} .
\end{align}
Choosing
\begin{align}
    \ell_1 = \frac{\xi_s}{\xi^* + \xi_s + \xi_t} \, d_\mathrm{sep} , \qquad
    \ell_2 = \frac{\xi_t}{\xi^* + \xi_s + \xi_t} \, d_\mathrm{sep} ,
\end{align}
and defining
\begin{align}
    \xi'' = \xi^* + \xi_s + \xi_t ,
\end{align}
we arrive at
\begin{align}
    &\| \Tr_C [\rho_C^{(0)} \, V_{BC}^{\mathrm{I}}(s) \, \bar{V}_{BC}(\beta-t,-t)]
    - \Tr_C[\rho_C^{(0)} \, V_{BC}^{\mathrm{I}}(s)] \,
      \Tr_C[\rho_C^{(0)} \, \bar{V}_{BC}(\beta-t,-t)] \| \notag\\
    &\le
    \mathrm{const.} \times |\partial B| \, |\partial B'| \,
    e^{-d_\mathrm{sep} / \xi''} .
\end{align}
Therefore, the first term in Eq.~\eqref{eq:S1_double_boundary_rewrite} is bounded by
\begin{align}
    &\Bigl| \Tr_{AB}[\tilde{\rho}_{AB}
    \{ \Tr_C [\rho_C^{(0)} \, V_{BC}^{\mathrm{I}}(s) \, \bar{V}_{BC}(\beta-t,-t)]
    - \Tr_C[\rho_C^{(0)} \, V_{BC}^{\mathrm{I}}(s)] \,
      \Tr_C[\rho_C^{(0)} \, \bar{V}_{BC}(\beta-t,-t)] \}
    \, o_A(\beta)] \Bigr| \notag\\
    &\le
    \mathrm{const.} \times |\partial B| \, |\partial B'| \, \|o_A\| \,
    e^{-d_\mathrm{sep} / \xi''} .
\end{align}
The second term in Eq.~\eqref{eq:S1_double_boundary_rewrite} is treated in the
same way. Hence the contribution containing $\Delta_{BB'}(s,t)$ is also
exponentially suppressed in $d_\mathrm{sep}$.

Collecting the bounds on the $\Delta(s,t)$, $\Delta_B(s,t)$, $\Delta_{B'}(s,t)$,
and $\Delta_{BB'}(s,t)$ contributions, we conclude that each second-order term
in Eq.~\eqref{eq:S1_diff_VAP_LVP} decays exponentially with $d_\mathrm{sep}$. Therefore,
\begin{align}
    |\langle o_A \rangle_\mathrm{VAP}
    - \langle o_A \rangle_\mathrm{LVP}|
    \le
    \mathrm{const.} \times \lambda^2 \, |\partial B| \, |\partial B'| \, \|o_A\| \,
    e^{-d_\mathrm{sep} / \xi_{\mathrm{eff}}}
    + \mathcal{O}(\lambda^3) ,
\end{align}
for some effective correlation length $\xi_{\mathrm{eff}}$.
This proves Proposition~\ref{prop:VAP-LVP}.

\end{document}